\documentstyle[epsfig]{mn}

\newif\ifAMStwofonts

\ifoldfss
  \ifCUPmtlplainloaded \else
    \NewTextAlphabet{textbfit} {cmbxti10} {}
    \NewTextAlphabet{textbfss} {cmssbx10} {}
    \NewMathAlphabet{mathbfit} {cmbxti10} {} 
    \NewMathAlphabet{mathbfss} {cmssbx10} {} 
  \fi
  \ifAMStwofonts
    \ifCUPmtlplainloaded \else
      \NewSymbolFont{upmath} {eurm10}
      \NewSymbolFont{AMSa} {msam10}
      \NewMathSymbol{\upi}     {0}{upmath}{19}
      \NewMathSymbol{\umu}     {0}{upmath}{16}
      \NewMathSymbol{\upartial}{0}{upmath}{40}
      \NewMathSymbol{\leqslant}{3}{AMSa}{36}
      \NewMathSymbol{\geqslant}{3}{AMSa}{3E}

    \fi
  \fi
\fi 

\ifnfssone
  \newmathalphabet{\mathit}
  \addtoversion{normal}{\mathit}{cmr}{m}{it}
  \addtoversion{bold}{\mathit}{cmr}{bx}{it}
  \newmathalphabet{\mathbfit} 
  \addtoversion{normal}{\mathbfit}{cmr}{bx}{it}
  \addtoversion{bold}{\mathbfit}{cmr}{bx}{it}
  \newmathalphabet{\mathbfss} 
  \addtoversion{normal}{\mathbfss}{cmss}{bx}{n}
  \addtoversion{bold}{\mathbfss}{cmss}{bx}{n}
  \ifAMStwofonts
    \ifCUPmtlplainloaded \else
      \UseAMStwoboldmath
      \makeatletter
      \new@mathgroup\upmath@group
      \define@mathgroup\mv@normal\upmath@group{eur}{m}{n}
      \define@mathgroup\mv@bold\upmath@group{eur}{b}{n}
      \edef\UPM{\hexnumber\upmath@group}
      \new@mathgroup\amsa@group
      \define@mathgroup\mv@normal\amsa@group{msa}{m}{n}
      \define@mathgroup\mv@bold\amsa@group{msa}{m}{n}
      \edef\AMSa{\hexnumber\amsa@group}
      \makeatother
      \mathchardef\upi="0\UPM19
      \mathchardef\umu="0\UPM16
      \mathchardef\upartial="0\UPM40
      \mathchardef\leqslant="3\AMSa36
      \mathchardef\geqslant="3\AMSa3E
    \fi
  \fi
\fi 

\ifnfsstwo
  \DeclareMathAlphabet{\mathbfit}{OT1}{cmr}{bx}{it}
  \SetMathAlphabet\mathbfit{bold}{OT1}{cmr}{bx}{it}
  \DeclareMathAlphabet{\mathbfss}{OT1}{cmss}{bx}{n}
  \SetMathAlphabet\mathbfss{bold}{OT1}{cmss}{bx}{n}
  \ifAMStwofonts
    \ifCUPmtlplainloaded \else
      \DeclareSymbolFont{UPM}{U}{eur}{m}{n}
      \SetSymbolFont{UPM}{bold}{U}{eur}{b}{n}
      \DeclareSymbolFont{AMSa}{U}{msa}{m}{n}
      \DeclareMathSymbol{\upi}{0}{UPM}{"19}
      \DeclareMathSymbol{\umu}{0}{UPM}{"16}
      \DeclareMathSymbol{\upartial}{0}{UPM}{"40}
      \DeclareMathSymbol{\leqslant}{3}{AMSa}{"36}
      \DeclareMathSymbol{\geqslant}{3}{AMSa}{"3E}
    \fi
  \fi
\fi 

\ifCUPmtlplainloaded \else
  \ifAMStwofonts \else 
    \def\upi{\pi}
    \def\umu{\mu}
    \def\upartial{\partial}
  \fi
\fi

\newcommand{\ea}{et~al.\ }
\newcommand{\ppi}{Paper I}
\newcommand{\ppii}{Paper II}
\newcommand{\pr}{\textrm{Pr}}
\newcommand{\real}{\textrm{Re}}
\newcommand{\imag}{\textrm{Im}}

\title
[Maximum-entropy weak lens reconstruction]
{Maximum-entropy weak lens reconstruction: improved methods and
application to data}

\author[P.J.~Marshall \ea]
{P.J.~Marshall,$^1$\thanks{Send offprint requests to:
P.Marshall@mrao.cam.ac.uk} M.P.~Hobson,$^1$ S.F.~Gull,$^1$ S.L.~Bridle$^2$ \\
$1$ Astrophysics Group, Cavendish Laboratory, Madingley Road, Cambridge
CB3 0HE
\\
$2$ Institute of Astronomy, Madingley Road, Cambridge CB3 0HA 
\\ 
}

\date{Accepted ???. Received ???; in original form \today}

\begin{document}
%
%
\maketitle
%
%
\begin{abstract} 
We develop the maximum-entropy weak shear mass reconstruction
method presented in earlier papers by taking each 
background galaxy image shape as an independent estimator of the
reduced shear field and
incorporating an intrinsic smoothness
into the reconstruction. The characteristic length scale of this
smoothing is determined by Bayesian methods.
Within this algorithm the uncertainties due to both the intrinsic
distribution of galaxy shapes and galaxy shape estimation are carried 
through to the final
mass reconstruction, and the mass within arbitrarily shaped apertures
can be calculated with corresponding uncertainties.
We apply this method to two clusters taken from $n$-body simulations 
using
mock observations corresponding to Keck LRIS and mosaiced 
HST WFPC2 fields.
We demonstrate that the Bayesian choice of smoothing length is sensible
and that
masses within apertures (including one on a filamentary structure) are
reliable, provided the field of view is not too small.
We apply the method to data taken on the cluster MS1054-03
using the Keck LRIS
(Clowe et al. 2000) and HST (Hoekstra et al. 2000), finding results in
agreement with this previous work; we also present reconstructions with
optimal smoothing lengths, and mass estimates which do not rely on any
assumptions of circular symmetry.
The code used in this work (LensEnt2) is available from the web.
\end{abstract}



\section{Introduction}
\label{intro}

Weak lensing studies of clusters of galaxies are an important complement
to X-ray, Sunyaev-Zel'dovich effect and optical observations, allowing
the projected distribution of mass to be investigated without any 
dynamical
assumptions. The reconstruction of cluster mass distributions from weak
gravitational lensing data is now well established; it has been
shown that the projected density distribution can be recovered from 
magnification data, in the form of background galaxy number
densities~\cite{BTP95,DT98}, or from shear data, the net statistical
distortion of the images of background
galaxies~\cite{TVW90,KS93,SS95,SK96}. 

We focus here on shear data, primarily because of
its greater abundance; the likelihood function for shear data is
also better understood~(Section~\ref{method}). 
Schneider, King and Erben~\shortcite{SKE00} 
discuss the use of the two types of weak gravitational lensing 
data.

Reconstruction methods using shear data fall into two classes:
direct and iterative inverse methods. The direct methods are based on 
the
pioneering work of Kaiser and Squires~(1993, KS93); many improvements 
have since
been made to the original 
algorithm~\cite{SS95,Kai95,Bar95,SK96}. In all these methods the 
galaxy shape data have to be smoothed
before their input to the algorithm; the smoothing length is a 
parameter that is left undetermined. The class of iterative methods aims 
to find the mass or projected
gravitational potential map that best fits the data~\cite{SK96,BNSS,SSB98}.
These methods are well suited to irregularly shaped observations, since
they do not suffer from edge effects in the same way as the direct
methods; however, they
need to be regularised in some way to prevent over-fitting the data,
and it remains unclear how best to determine the resolution of either
the data bins or the reconstruction grid.

In two earlier papers~(Bridle~\ea 1998, \ppi; Bridle~\ea 2001, \ppii)
we presented a maximum-entropy
inverse
method for reconstructing the mass distribution in clusters using shear
and/or magnification data. In this paper we extend our method to give a
fuller Bayesian analysis. As noted by other authors~\cite{SSB98}, 
it would be
desirable to work with each background galaxy shape individually, rather
than binning or smoothing the data. This issue, together with the
problem of the angular
resolution of the reconstruction, is addressed by our extended
algorithm.  We apply our improved
method to both realistic synthetic data, and previously published data
for the high-redshift cluster MS1054-03. As with any Bayesian analysis,
the aim is to derive and interpret the
full posterior probability
distribution of the quantity being inferred (in this case the mass
distribution and any associated parameters). 
This approach will provide us not just with a mapping procedure, but
also valuable insight into the quality of the data itself. 

The method is reviewed and further developed in Section~\ref{method},
and is applied to simulated data in Section~\ref{simobs}.
Section~\ref{realdata} contains the results of our method applied to the
well-documented cluster MS1054-03, and gives a brief comparison 
with the
previously published work. Our conclusions are presented in
Section~\ref{conc}.


\section{Method}
\label{method}

The basis of the weak lensing reconstruction method described here is 
essentially that of \ppi; this section presents several 
developments in the algorithm and its implementation.

A trial mass distribution $\Sigma(\btheta)$ is
used to generate a
predicted reduced shear field $g(\btheta)$ through the
convolution~(Kaiser \& Squires 1993, \ppi) 
\begin{equation}
g(\btheta) = \frac{1}{1-\kappa(\btheta)} \cdot \frac{1}{\pi} 
\int D(\btheta-\btheta') \kappa(\btheta') d^2 \btheta',
\label{eq:k2g}
\end{equation}
where the convergence 
$\kappa(\btheta) = \Sigma(\btheta) / \Sigma_{\rm crit}(\btheta)$
and $\Sigma_{\rm crit}$ is a factor dependent on the lens and source
redshifts. 

By design the lensing convolution kernel~$D$ is a complex quantity that
picks out the two 
types of lensing distortion $g_1=\real(g)$ and $g_2=\imag(g)$.
Unbiased estimates of these
components of reduced shear are given by the ensemble
average of the  background
galaxy image ellipticity parameters $\epsilon_1$ and
$\epsilon_2$~\cite{SK95}. 

As in Papers I and II, we aim to reconstruct the projected mass density of the 
lens defined on a 
grid of square pixels, where the observing region occupies a
smaller area within this
grid. This allows for the fact that the mass outside the observed field
affects the shear data inside. 
It has been noted~\cite{SSB98} that reconstructing the projected
lensing potential allows a purely local estimate of the mass
distribution to be derived, by numerical differentiation of the 
potential. This last step involves throwing away a small amount of
information, that which describes the mass distribution outside the
observing field. Although, as Seitz~\ea point out, this information is
limited, we feel it is as well to try and include it for completeness.
In most cases, the cluster being studied will lie completely within the
observing field and the two reconstruction approaches should produce
indistiguishable results; it is then a matter of taste as to which
quanitity is inferred. Since here we are interested in the
masses of clusters, we choose to reconstruct the surface mass density
directly, leading to simply-estimated projected masses with 
well-understood derived uncertainties. 

\subsection{Using individual galaxy shapes}

In \ppi, the predicted reduced shear was 
compared with measured
galaxy ellipticities averaged in coarse grid cells. Following
Seitz \ea\shortcite{SSB98}, we prefer to use each galaxy shape
individually, as 
independent estimators of the reduced shear. 
This procedure removes
the potential problem of the bin boundaries affecting the 
inferred mass distribution,
allows for optimal angular resolution in the
reconstruction, and leaves the data in as pure a form as possible.
The reconstruction grid
pixel size is chosen to have approximately 1 galaxy per pixel,
leading to comparable numbers of data points and fitted parameters. 
However, each data point has a very low signal-to-noise ratio,
indicating that the number of parameters should be reduced in some
way -- this issue is addressed in the next section.
The convolution of Eq.~(\ref{eq:k2g}) is performed using Fast Fourier
Transforms, and the resulting reduced shear field is
interpolated onto the background galaxy positions. 

Each of the~$2N$
lensed ellipticity components~$\epsilon_j$ of the~$N$ measured background
galaxy images are taken as having been
drawn independently from
a Gaussian distribution with mean~$g_j$ and 
variance~$\sigma_{\rm intrinsic}^2$;
here~$g_j$ is the true value of the $j^{\rm th}$ component of 
reduced shear at the position of the
galaxy.
We can then
write the likelihood function as 
\begin{equation}
\pr(\mbox{\mbox{Data}}|\Sigma) = \frac{1}{Z_L} \exp ( -\frac{\chi^2}{2} )
\label{eq:lhood}
\end{equation} 
where $\chi^2$ is the usual misfit statistic 
\begin{equation}
\chi^2 = \sum_{i=1}^{N} \sum_{j=1}^{2} \frac{(\epsilon_{j,i} -
g_{j,i})^2}{\sigma^2},
\label{eq:chisq}
\end{equation} 
and the normalisation factor is
\begin{equation}
Z_L = (2 \pi \sigma^2)^{\frac{2N}{2}}
\label{eq:chisqnorm}
\end{equation} 

The effect of errors introduced by the galaxy shape estimation procedure
have been included by adding them in quadrature to the intrinsic
elipticity dispersion~\cite{HFK00}, 
\begin{equation}
\sigma = \sqrt{\sigma^2_{\rm obs}+\sigma^2_{\rm intrinsic}}
\label{eq:newsigma}
\end{equation} 
This approximation
rests on the assumption
that 
both the shape estimation error and
the unlensed ellipticity distributions 
are fitted well by Gaussians,
and that the applied reduced shear is not too large.
We follow
Schneider~\ea\shortcite{SKE00} and 
correct the
width of the ellipticity distributions by a factor of $(1-|g|^2)$ to
account for the non-linearity in the lensing 
transformation~(equation~\ref{eq:lens} below). 
We are concerned here with sub-critical clusters for which this
correction factor is small; in principle, the likelihood may be refined
to include other effects as well. 
In practice, we find that this particular correction makes little 
difference to the reconstructions.

\subsection{The ICF and Bayesian evidence}

Our inferences of the distribution of $\Sigma$ in the cluster are based 
on the posterior probability distribution given by Bayes' theorem:
\begin{equation}
\pr(\Sigma|\mbox{Data}) =
\frac{\pr(\mbox{Data}|\Sigma)\pr(\Sigma)}{\pr(\mbox{Data})}.
\label{eq:bayes1}
\end{equation} 
In \ppi, an entropic prior $\pr(\Sigma)$ was introduced for this 
positive additive
distribution; maximisation of 
$\pr(\Sigma|\mbox{Data})$ then reduces to minimisation of the
function $F = \chi^2/2 - \alpha S$, where~$S$ is the entropy function
for the distribution.
At this point the method is essentially an entropy-regularised
maximum-likelihood technique, similar to that published elsewhere by 
Seitz, Schneider and Bartelmann~\shortcite{SSB98}.  

This approach contains an implicit assumption that the values of 
$\Sigma$ are uncorrelated. However, we expect clusters of galaxies to
have smooth, extended projected mass distributions, and wish to
include this knowledge in our analysis. The Intrinsic Correlation
Function~(ICF) formalism \cite{Gull89,Rob92} allows us to do exactly
that; the physical distribution $\Sigma$ is expressed as the
convolution of a `hidden' distribution with a broad kernel (the
ICF). In this way the smoothing, which
is always necessary at some stage when using such noisy data, is
transferred from the data to the reconstruction process itself.
The large number of free parameters in the model 
(the hidden pixel values) is
effectively reduced by this smoothing to a number appropriate to the
quality of the data.
In this way the properties of the noise can be carried through in a 
calculable, if
non-linear, fashion. Seitz~\ea\shortcite{SSB98} incorporate smoothing in
their reconstruction scheme, but in an iterative way, making error
estimation non-trivial.

We now consider the form of the ICF; its parameterisation
introduces new degrees of freedom into the problem. 
A Bayesian analysis should
allow the data to dictate the most suitable ICF, as follows.
We expect the most important parameter of the ICF to be its width.
For a given functional form (e.g. a circularly-symmetric Gaussian), 
depending on a single
width parameter $w$, 
equation~(\ref{eq:bayes1}) reads
\begin{equation}
\pr(\Sigma|\mbox{Data},w) =
\frac{\pr(\mbox{Data}|\Sigma,w)\pr(\Sigma|w)}{\pr(\mbox{Data}|w)}.
\label{eq:bayes2}
\end{equation}
The width parameter could be chosen to maximise its own posterior
probability~$\pr(w|\mbox{Data})$; this distribution would certainly be a
useful tool in
assessing the relative merits of different ICF widths. 
Bayes' theorem again gives 
\begin{equation}
\pr(w|\mbox{Data}) = \frac{\pr(\mbox{Data}|w)\pr(w)}{\pr(\mbox{Data})}.
\label{eq:bayes3}
\end{equation}
Usually, the typical angular scale of the cluster is known to
within at least an order of magnitude so that a uniform prior for $w$ is
appropriate; since $\pr(\mbox{Data})$ is a constant, 
$\pr(\mbox{Data}|w)$ may be
used directly to infer $w$. This value can be obtained during the
reconstruction process by numerically evaluating the normalising factor in
equation~\ref{eq:bayes2},
\begin{equation}
\pr(\mbox{Data}|w) = \int \pr(\mbox{Data}|\Sigma,w)\pr(\Sigma|w)
d\Sigma.
\label{eq:marginalise}
\end{equation}
This integral is known as the `evidence'; Sivia~(1996) and MacKay~(1992)
give explanation of its application in data analysis. The evidence
provides an objective
discriminator between ICF widths~$w$, and, indeed, any other 
parameters we might choose to include in the reconstruction process.
Comparison of the evidence calculated for different functional forms of the
ICF allows the merits of different smoothing kernels to be
evaluated~(see section~\ref{advanced} below). 
Indeed, 
the regularisation parameter~$\alpha$~(\ppi) 
is also determined 
by maximising the evidence with respect to $\alpha$~\cite{Gull89}. 
Parameters such as~$\alpha$ and $w$ may be viewed as `nuisance'
parameters, and marginalised over. When the evidence 
is sharply
peaked at some value, this marginalisation is approximately equivalent 
to using the peak value~\cite{Mac92}.

Interpolation of a fine grid of predicted reduced shear values onto 
the galaxy
positions retains the potential to obtain high angular resolution
reconstructions; inclusion of an ICF effectively reduces the number
of independent pixels to one more appropriate to the quality of 
the data
at hand. An increase in the number of pixels in the working grids and
the inclusion of 
an extra convolution calls for a faster numerical algorithm than that
used in Papers~I and~II; we have utilised the commercially available
software MEMSYS4, developed by MaxEnt Data Consultants Ltd. This code is
widely used in the image processing community and has been proven to be
highly stable; details of the numerical algorithms can be found in 
Gull~\shortcite{Gull90}. 

\subsection{Quantitative mass mapping}

A side-effect of smoothing data prior to an 
inversion, or of incorporating an
intrinsic correlation function as described above,
is the introduction of
pronounced correlations between the errors on each reconstruction 
pixel value.
However,
calculation of the (Gaussian approximation to the) full covariance matrix 
of the errors on the $\Sigma$
distribution~(\ppi) successfully accounts for these correlations when
calculating integrals over the reconstruction. 
This is of particular
interest for the direct estimation of the total projected mass within 
an aperture from the reconstruction map. Information on the 
shape of the aperture is contained within a vector of weights~$c_i$.
The constant 
$c_i$ is equal to zero if the $i^{\rm th}$~pixel lies completely outside the
aperture; otherwise $c_i = A_i$, where~$A_i$ is the area of the 
$i^{\rm th}$~pixel (in
square parsecs at the cluster) lying within the aperture. We then
approximate the integral by a weighted sum of pixel values to produce
the mass estimate 
$(M \pm \sigma_M)$, where
\begin{equation}
M = \sum_i c_i \Sigma_i 
\label{eq:massest}
\end{equation}
and
\begin{equation}
\sigma_M = \sum_{ij} c_i c_j V_{ij}.
\label{eq:massesterr}
\end{equation}
Here $V_{ij}$ is the covariance matrix of the reconstruction errors in
each pixel~(\ppi).
Whilst a parameterised fit~\cite{SKE00,KS01}
may be more physically motivated, this mass estimation
procedure provides a quantitative result that can be used to guide
further analysis.  
The  
aperture can be any shape, and so can be tailored to match the
investigation in hand. Also, calculation of a realistic error on the
mass estimates allows, within the Gaussian approximation, estimation of
the significance of features in the maps, without recourse 
to the peaks analysis or resampling methods advocated 
elsewhere~\cite{vW00,E++00,Hoe++01}.

\begin{figure*}
\begin{minipage}[t]{0.48\linewidth}
\centering\epsfig{file=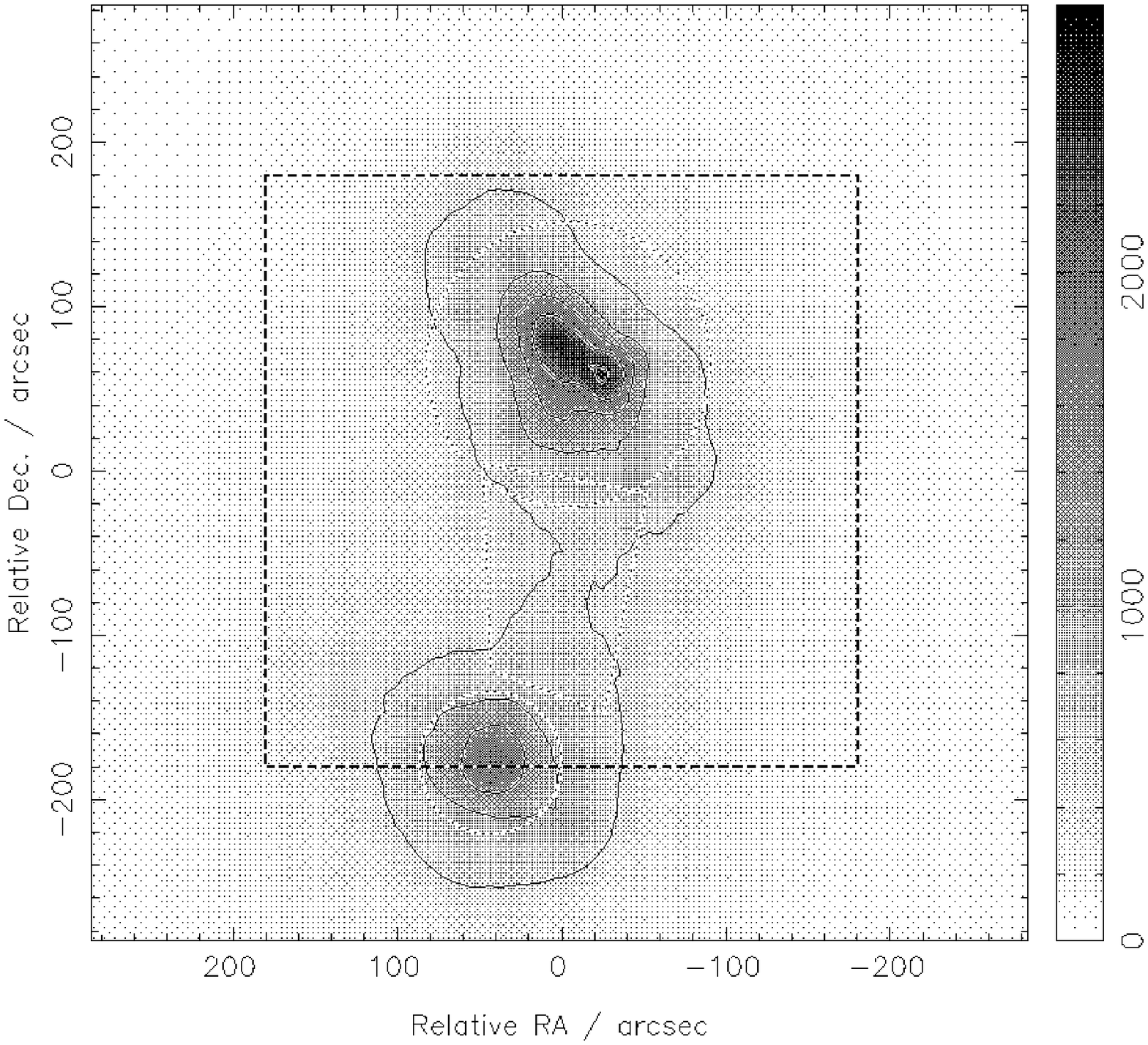,width=\linewidth,angle=0,clip=}
\end{minipage} \hfill
\begin{minipage}[t]{0.48\linewidth}
\centering\epsfig{file=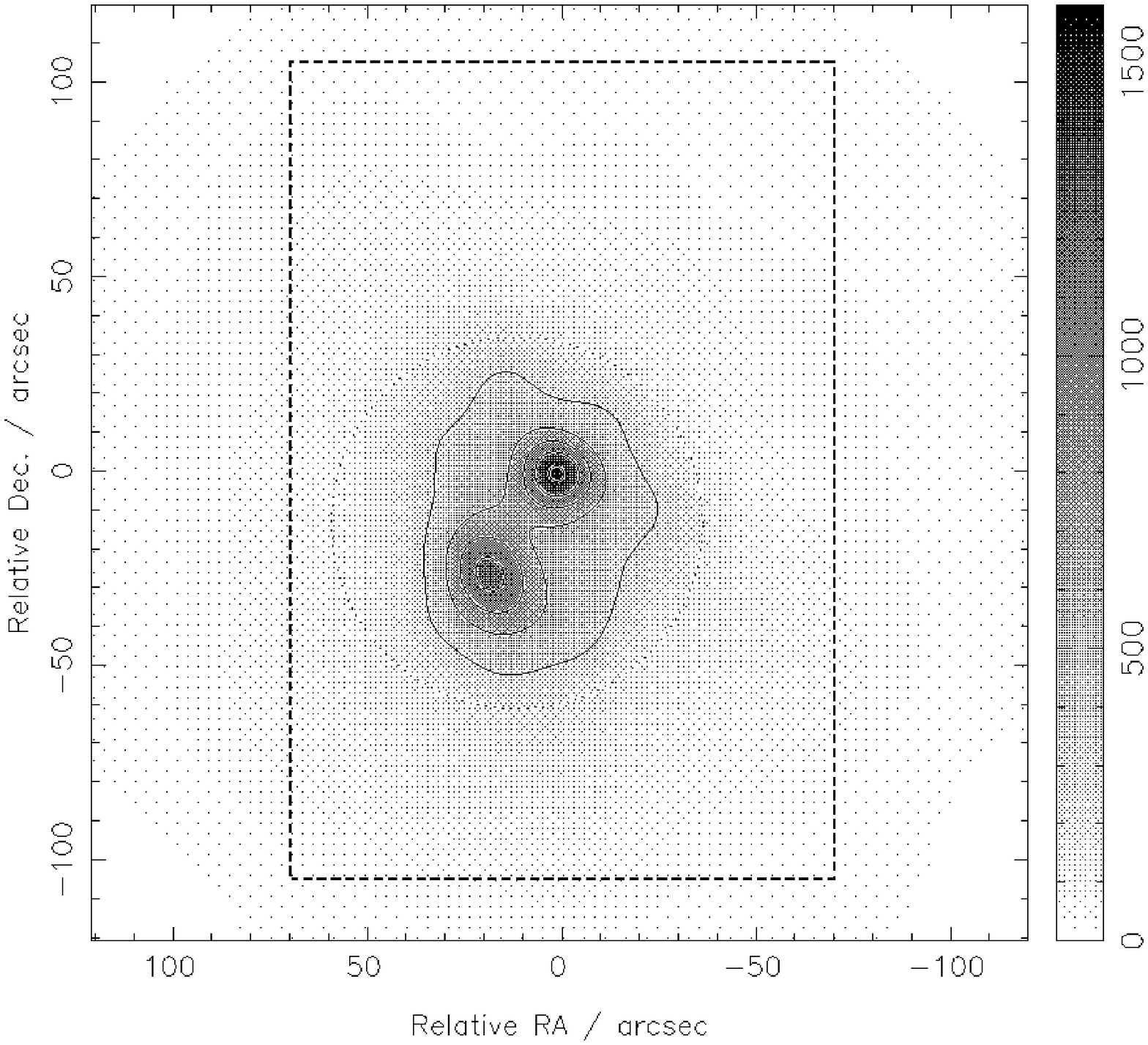,width=\linewidth,angle=0,clip=}
\end{minipage}
\caption{Projected mass distribution of two simulated clusters. Left:
CL10, a massive cluster at redshift 0.2. Right: CL08, a smaller cluster
at redshift 0.78. The grey scale is~$\Sigma$/M$_{\odot}$\,pc$^{-2}$,
the contours are spaced in steps of 500~h~$M_{\odot}$ pc$^{-2})$ 
for CL10 and 300~h~$M_{\odot}$ pc$^{-2})$ for CL08.
Also marked are the apertures used for mass estimation from the
reconstructed maps (dotted) and the mock observing region (dashed).}
\label{fig:cltrue}
\end{figure*}


\section{Application to simulated data}
\label{simobs}

To demonstrate the method outlined in the previous section we
now apply it to two simulated clusters, taken from the sample
generated by Eke, Navarro and Frenk~\shortcite{ENF98}. The naming of
the clusters is retained from that paper, in which a flat Universe
dominated by a cosmological constant was assumed. The same cosmological
parameters ($\Omega_m=0.3,\Omega_{\Lambda}=0.7$) were used to calculate
the critical density and angular diameter distances needed in the
lensing analysis below. The Hubble constant is taken as 
$100 h$~km~s$^{-1}$~Mpc$^{-1}$.

\subsection{The mass distributions}
\label{massdist}

The two projected mass
distributions are shown in Figure~\ref{fig:cltrue}. The first, CL10, is
at a redshift of~$0.2$, and has an X-ray emission-weighted temperature
of~$4.0$~keV; the second, CL08, is at redshift 0.78 with
temperature~$2.1$~keV. Neither cluster is extremely massive, having
approximate virial masses of $6$ and $1 \times 10^{14} h^{-1} M_{\odot}$
respectively. Overlaid on these plots are the observing
regions corresponding approximately to the field of view of the Keck 
Low Resolution Imaging Spectrograph 
(for CL10, see e.g.~Clowe~\ea\shortcite{Clo++00}) and an HST mosaic
comprising two WFPC2 pointings (for CL08). 
Also plotted are the apertures
used to estimate projected masses for the cluster components.
For CL10, these are circles of radius 0.2~h$^{-1}$Mpc ($\sim 87$~arcsec)
and 0.1~h$^{-1}$Mpc ($\sim 43$~arcsec) centred on the large and small
subclumps respectively, and are referred to as apertures 1 and 2.
Aperture 3 is the quadrilateral region between the subclumps.
For CL08 a single aperture is defined, being a circle of radius 
0.25~h$^{-1}$Mpc ($\sim 48$~arcsec). 
Neither observing
region is particularly large, and the smaller clump in CL10 is
marginally outside the observing field. These clusters were deliberately
chosen to contain `interesting' substructure, in order to illustrate the
angular resolution of the reconstruction method.

\subsection{Simulated lensing data}
\label{simdata}

\begin{figure*}
\begin{minipage}[t]{0.48\linewidth}
\centering\epsfig{file=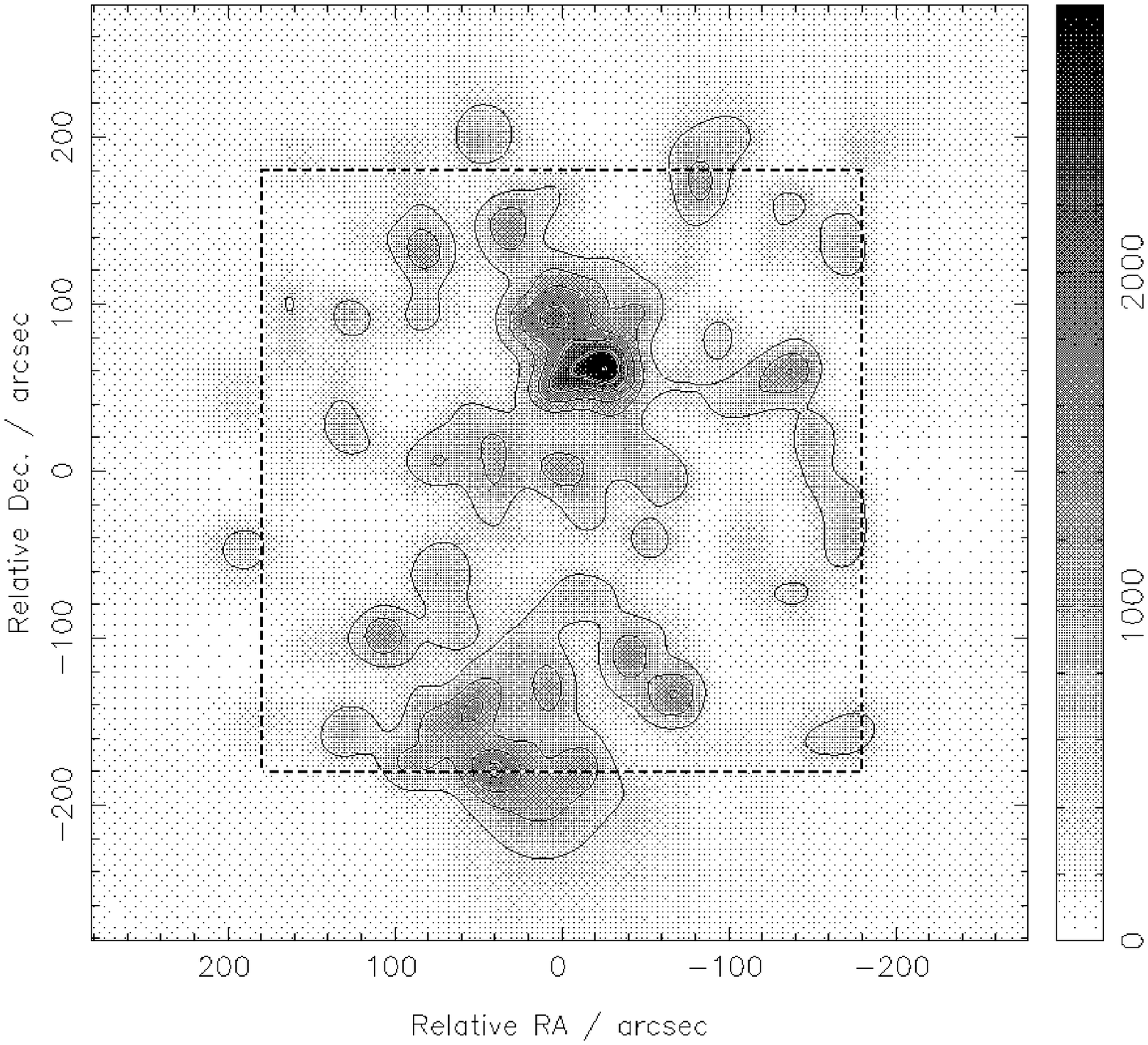,width=\linewidth,angle=0,clip=}
\end{minipage} \hfill
\begin{minipage}[t]{0.48\linewidth}
\centering\epsfig{file=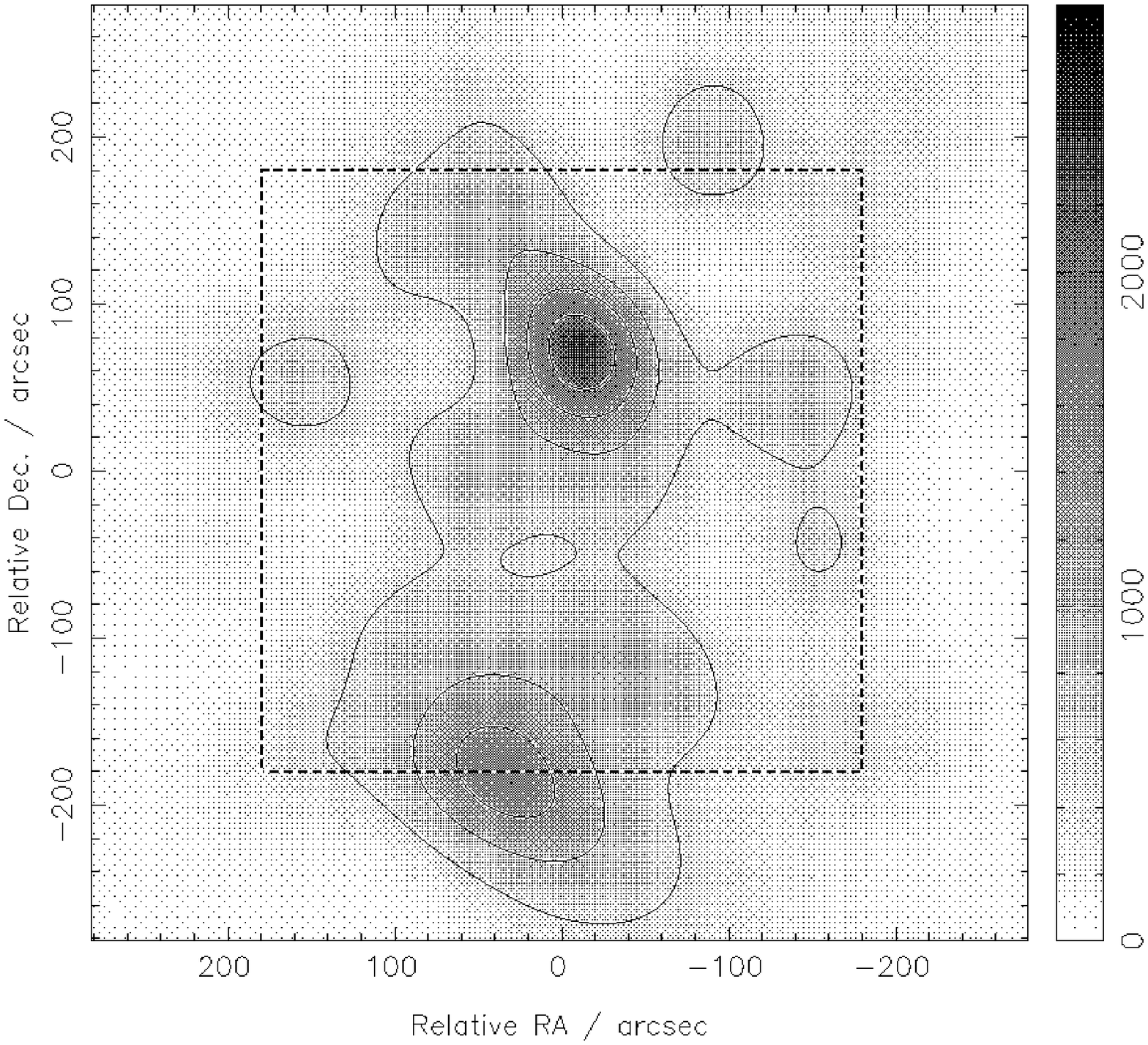,width=\linewidth,angle=0,clip=}
\end{minipage} 
\vspace{5mm}

\begin{minipage}[t]{0.48\linewidth}
\centering\epsfig{file=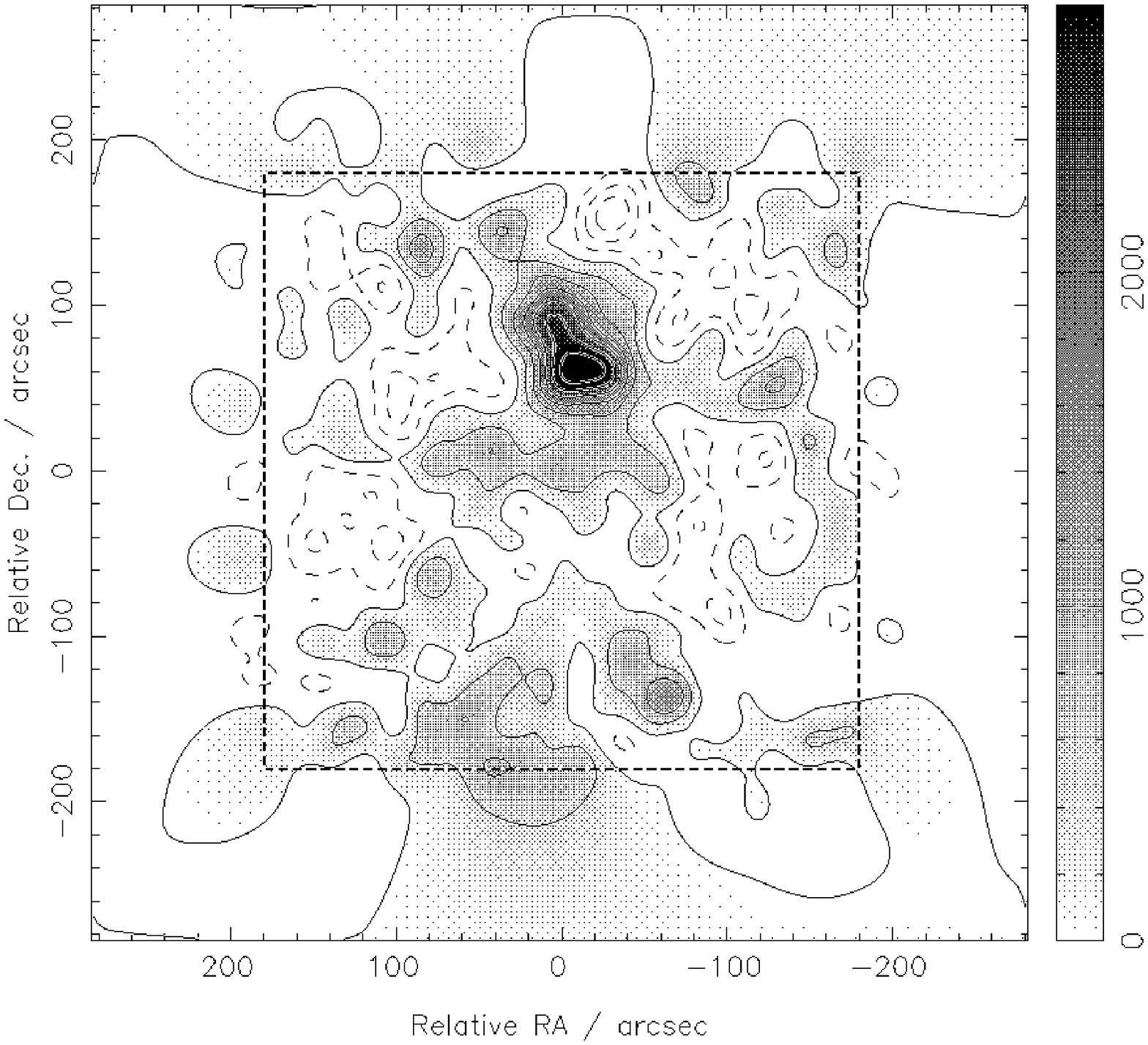,width=\linewidth,angle=0,clip=}
\end{minipage} \hfill
\begin{minipage}[t]{0.48\linewidth}
\centering\epsfig{file=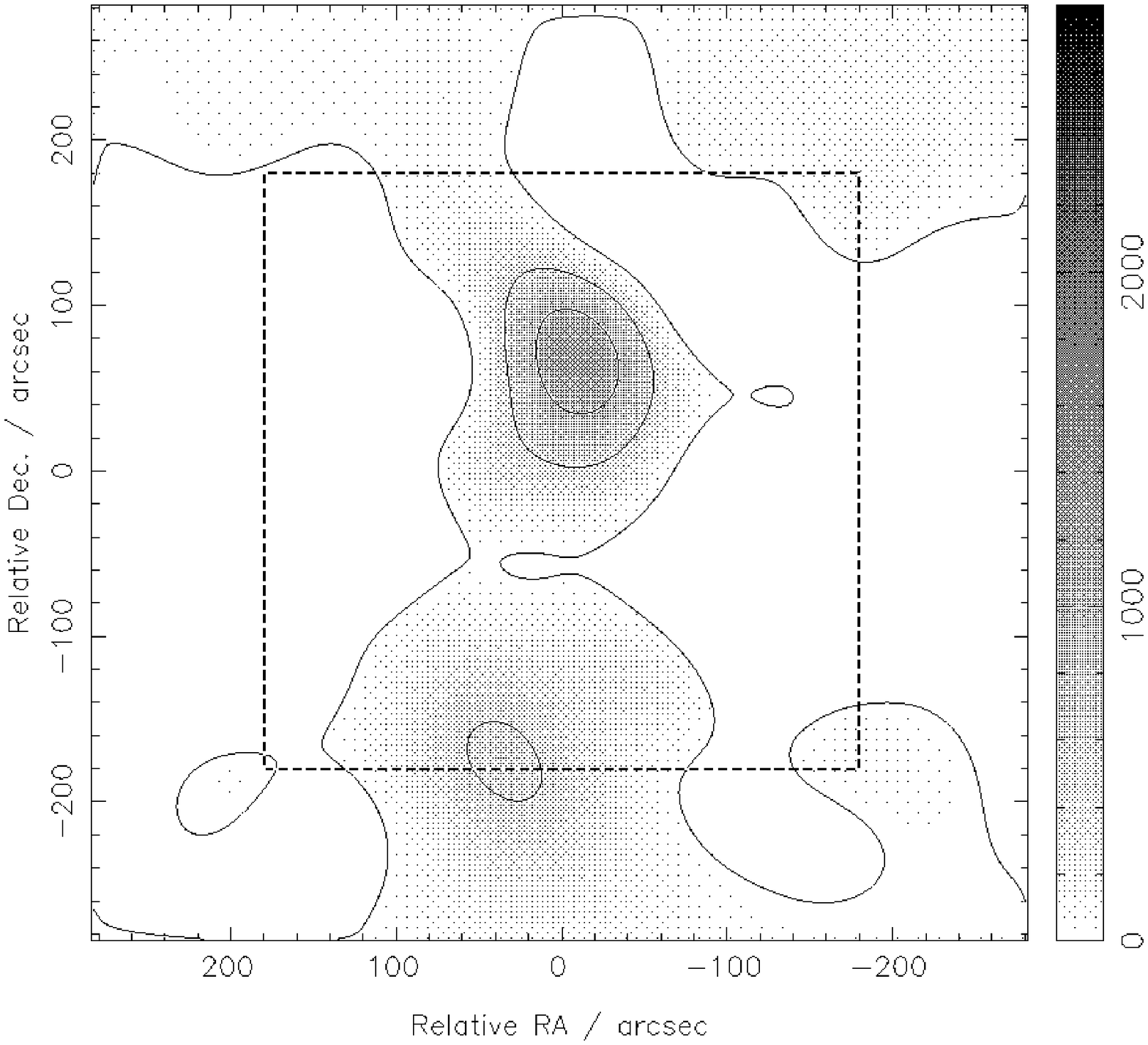,width=\linewidth,angle=0,clip=}
\end{minipage}
\caption{Top: Reconstructed mass distributions for the cluster CL10.
Left to right, the ICF width parameter~$w$ increases from 20 to
70~arcsec. 
Bottom: KS93 direct inversions for comparison; the shear data
were smoothed with a Gaussian of FWHM equal to~$w$ in each case. In all
plots the contours show surface density in steps of 500~h~$M_{\odot}$
pc$^{-2}$. The maximum on the density scale corresponds to a 
convergence of 0.61.}
\label{fig:cl10recon}
\end{figure*}

Mock galaxy ellipticity catalogues were generated using the CL10 and
CL08 mass distributions as follows. 
The parameters of the observations described in 
Clowe~\ea\shortcite{Clo++00} and Hoekstra~\ea\shortcite{HFK00} 
were used to estimate the background galaxy number 
densities obtainable in 2 hours observation with either the Keck LRIS or
HST. The median galaxy redshift was estimated from the Hubble Deep Field photometric
redshift catalogue~\cite{FLY99}, and then used to calculate an
approximate value of~$\Sigma_{\rm crit}$. 
Systematic effects due to the unknown redshift distribution of
background galaxies~\cite{F+T97} are not considered here. 
The projected mass
distributions of Figure~\ref{fig:cltrue} were then converted to
convergence and equation~(\ref{eq:k2g}) was applied to produce maps of 
the
components of reduced shear~$g$ on the same grid. These maps were then
interpolated onto galaxy positions drawn at random from within the
observing field. The components of the intrinsic ellipticity of the 
sources,~$\epsilon^s_i$, were then drawn from a Gaussian distribution of
width~$\sigma_{\rm intrinisic}=0.25$ and transformed to their lensed counterparts using the
relation~\cite{SS97}
\begin{equation}
\epsilon =  \frac{\epsilon^s + g}{1 + g^*\epsilon^s},
\label{eq:lens}
\end{equation}
where the asterisk denotes complex conjugation.
Uncertainties introduced in the estimation of galaxy shapes were
included by adding Gaussian noise with~$\sigma_{obs}=0.15$. This is a
reasonably
pessimistic approach, with only the fainter galaxies 
detected having errors of this size associated
with their ellipticities~\cite{HFK00,Bac++01,Bri++02}.  
The limiting magnitudes quoted should therefore be taken as those down 
to which image shape measurements could 
be made to this accuracy.
The parameters of the mock observations are given in
Table~\ref{tab:obs}.

\begin{table*}
\begin{minipage}{123mm}
\caption{Properties of the simulated 2 hour observations of
section~\ref{simobs}.}
\label{tab:obs}
\begin{tabular}{ccccccccc}
 Cluster & $z$  & $I_{\rm lim}$ & $z_s$ & $\Sigma_{\rm crit}$          & $\kappa_{\rm peak}$ &  $A_{\rm obs} $   &    $n_s$             & $N_s$ \ \\ \
         &      &               &       &  (h~$M_{\odot} \rm pc^{-2})$ &                     &  ($\rm arcmin^2$) &  ($\rm arcmin^{-2}$) &       \ \\ \
  CL10   & 0.2  &    26.3       &  1.0      &      4634     & 0.61 &         36    &      40    &     1476          \\
  CL08   & 0.78 &    26.5       &  1.3      &      4950     & 0.38 &         8.2   &      100   &     695           \\
\end{tabular}

\medskip
From left to right: cluster  name and redshift, limiting I
magnitude, median source redshift, corresponding value of the critical
density, cluster peak convergence, observing area,
number density of background galaxies, and the total number of sources
in the simulated catalogue.
\end{minipage}
\end{table*}

\subsection{Results}
\label{simresults}

Reconstructions were performed for a range of Gaussian ICFs with varying
FWHM~$w$; 
the mass distributions were defined on 128 by 128 pixel grids. Each
reconstruction, including aperture integrated mass estimation, 
required approximately 
two minutes of CPU time on a single R10000 processor of a Silicon
Graphics Origin 200. 

\subsubsection{CL10}
\label{cl10}

Reconstructions with ICF widths of 20 and
70~arcsec are shown in the top two panels of Figure~\ref{fig:cl10recon}. 
The greyscale is plotted with the same limits as used for
the true mass distribution of
Figure~\ref{fig:cltrue}. 
For comparison, we also show results for which 
the mock data was \emph{first} smoothed, using the same Gaussian ICFs as
the smoothing kernel,
and then inverted
directly using the Kaiser \& Squires~\shortcite{KS93} algorithm to
obtain a
convergence map; 
results are plotted in the lower panel of the
same figure. Here the contours are simply of the surface density,
obtained by
scaling the convergence by the relevant value of $\Sigma_{\rm crit}$; 
the greyscale is
again the same as Figure~\ref{fig:cltrue}. 
The grid on which the Fourier
transforms were performed was padded with zeros outside the observing
region to allow a more direct comparison with our method.  
At each smoothing scale the maps generated by the two methods 
contain recognisably similar
structures;
Figure~\ref{fig:cl10recon} illustrates the way in which the smoothing 
has been moved from the data to the reconstruction. 
Features which differ from the true mass distribution are similar in
both reconstructions, and are due to the noise realisation.

Compared to the KS93
results, the maximum-entropy
solutions are preferable as they are maps of inferred 
physical mass (so are necessarily positive), in which the noise
on each data point has been translated to inferred uncertainties in the
maps. At low values of the ICF width the
high noise in the shear data acts to break up the lensing signal,
leading to false apparent cluster substructure at small scales. The presence of 
many
low level spurious features is also due to this over-fitting of the
data. Note that the lensing signal is not diluted by moving to higher
values of~$w$, in contrast to the effect of
data smoothing in the KS93 process. 

\begin{figure}
\centering\epsfig{file=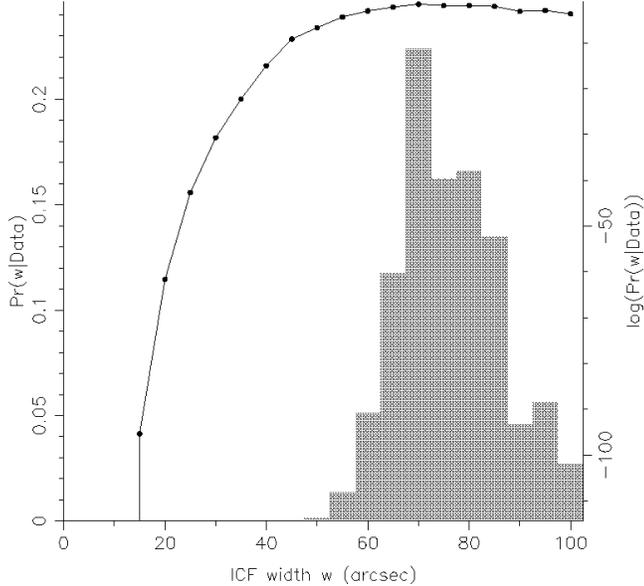,width=\linewidth,angle=0,clip=}
\caption{$\pr(w|\mbox{Data})$ for the CL10 analysis. The
logarithmic scale corresponds to the joined points, while the solid bars
are on the linear scale.}
\label{fig:cl10evid}
\end{figure}

\begin{figure}

\centering\epsfig{file=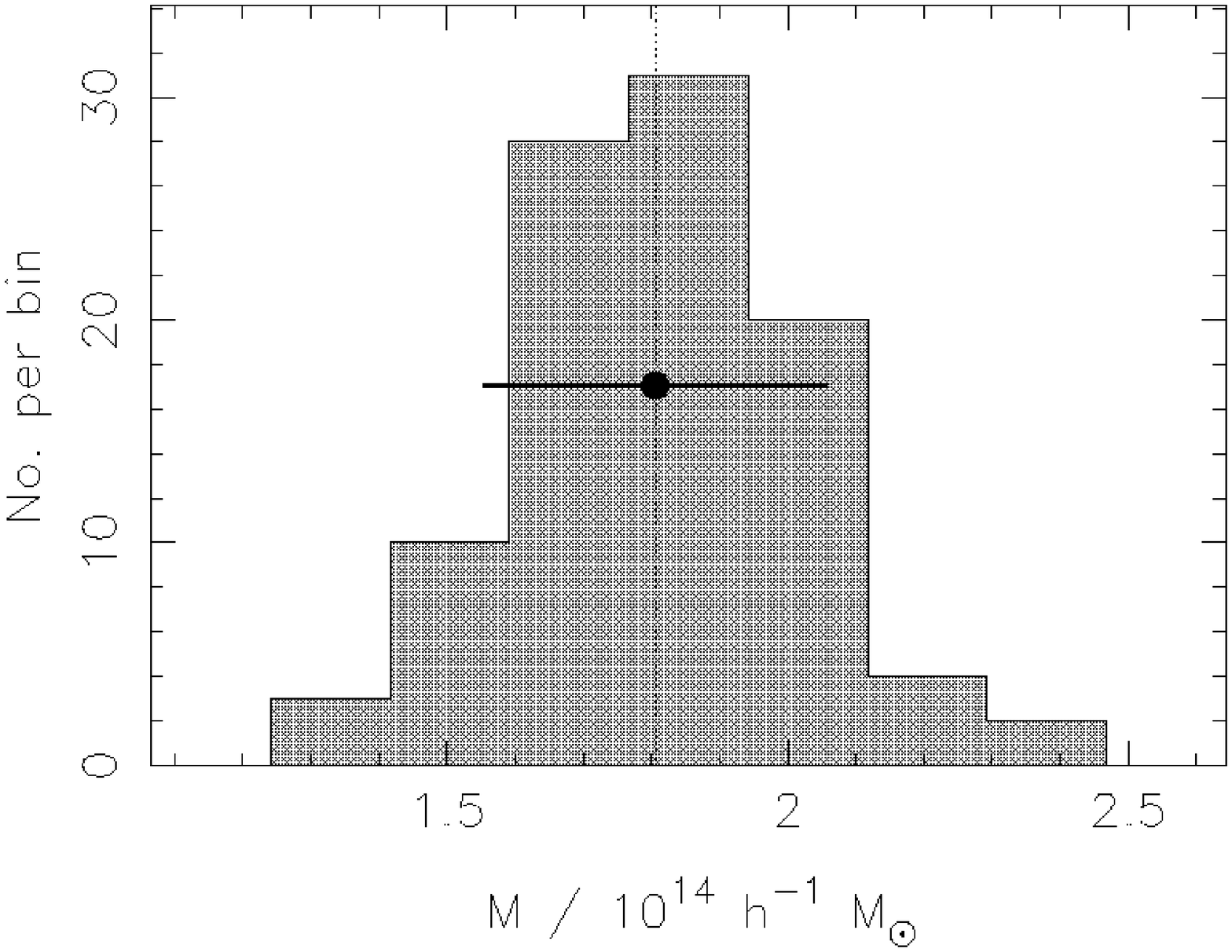,width=0.95\linewidth,angle=0,clip=}
\vspace{3mm}

\centering\epsfig{file=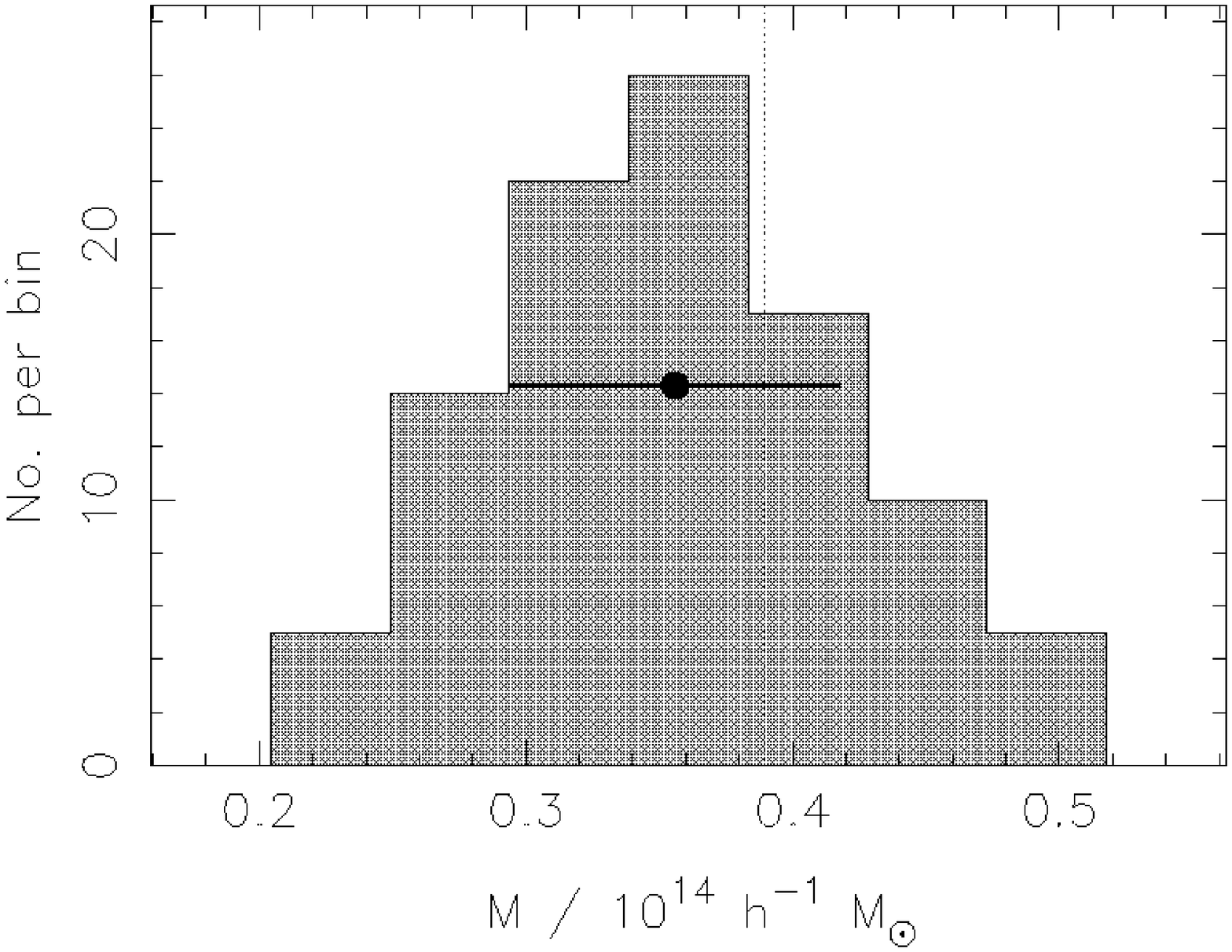,width=0.95\linewidth,angle=0,clip=}
\vspace{3mm}

\centering\epsfig{file=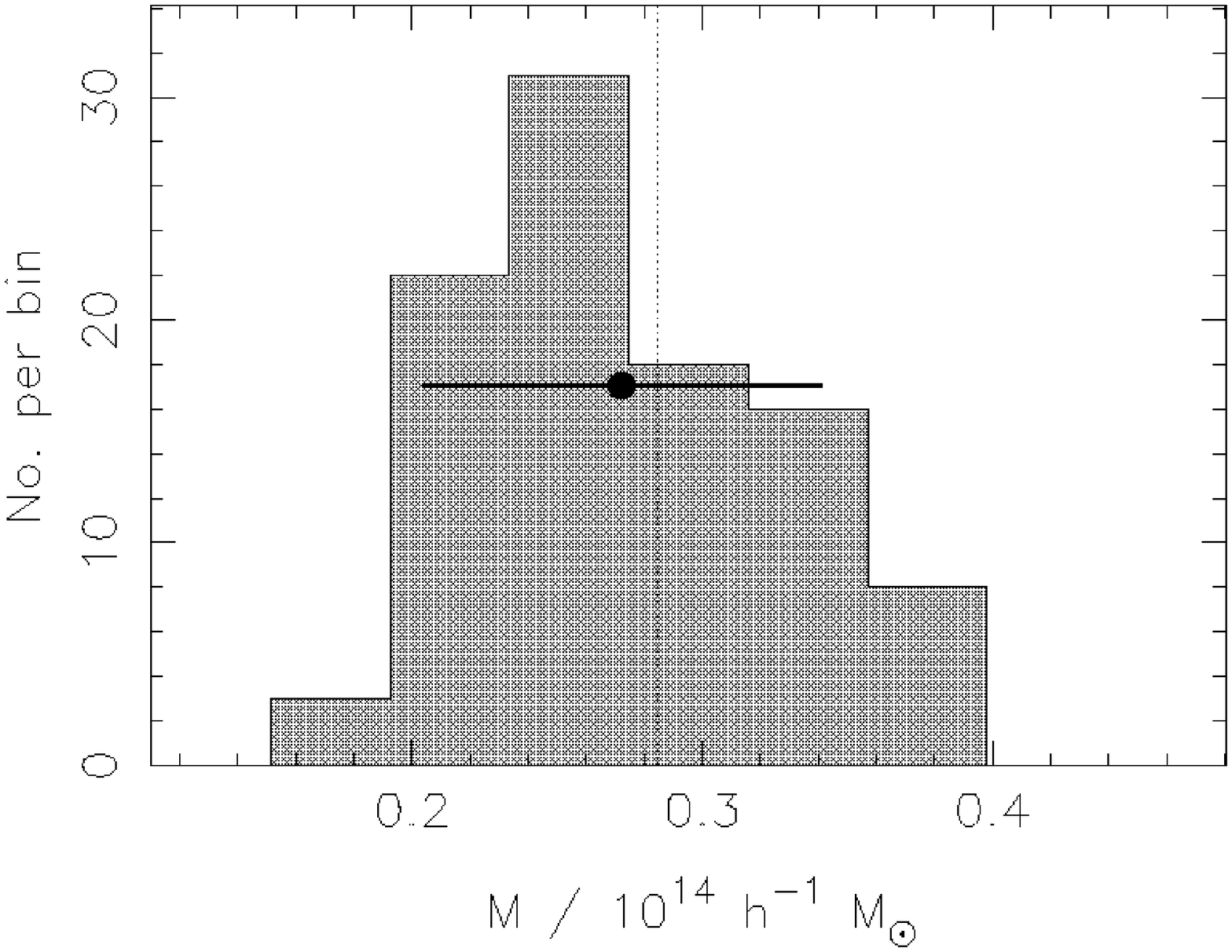,width=0.95\linewidth,angle=0,clip=}

\caption{Mass estimates for CL10. In each panel a histogram of mass
estimates from the reconstruction maps from 100 realisations of the
background galaxy population are plotted. The dotted line marks the true
value. The point shows the mean mass estimate; the error bar is the mean
inferred error, not the standard deviation of the histogram. The
apertures used are shown in Figure~\ref{fig:cltrue} Top:
circle of radius~$0.2 h^{-1}$ Mpc centred on the main cluster; middle:
circle of radius~$0.1 h^{-1}$ Mpc centred on the sub-clump; bottom:
quadrilateral region between the two main mass clumps.}
\label{fig:cl10hist}
\end{figure}

The two maximum-entropy
reconstructions in Figure~\ref{fig:cl10recon} are taken from the
posterior probability distribution~$\pr(w|\mbox{Data})$; as described in
Section~\ref{method} this distribution is proportional to the 
numerically evaluated evidence
and is
shown in Figure~\ref{fig:cl10evid}. The shape of this graph is typical. 
The reduction in
probability at large ICF widths is because the data are poorly fitted
by the overly smooth mass distribution.
At the other extreme, small ICF widths are strongly disfavoured as they 
effectively increase the number of free parameters in the fit; this is 
the `Occam's razor' factor which arises naturally from Bayesian model 
selection analysis~(e.g., MacKay 1992) and also 
corresponds to the intuition
that one shouldn't over- or under-smooth data.
The map corresponding to the maximum 
of~$\pr(w|\mbox{Data})$ represents a 
`map of believable features', and occurs at $w=70$~arcsec, shown in the
right-hand panel of Figure~\ref{fig:cl10recon}. 
The contours of this reconstruction
trace the two main mass condensations, and suggest the presence of a
bridge of mass between them; no other significant features are visible.
In the absence of further information about the cluster this map
represents the
most probable mass distribution, given the data and our choice of the
functional form of the ICF.
The
eye is very sensitive to the high
resolution detail of the main peak
in Figure~\ref{fig:cl10recon}; however, the 
evidence is sensitive to the entire mass
distribution, which is smooth on larger angular scales.  

The projected mass within each of the three apertures shown in
Figure~\ref{fig:cltrue} was calculated for the preferred 70 arcsec
ICF width. This process was carried out on reconstructions from 
100 galaxy catalogues
with the same observational parameters as in Table~\ref{tab:obs} 
but
with different galaxy positions and ellipticities.
The results are shown in the
histograms of Figure~\ref{fig:cl10hist}, with statistics from these 
measurements given in 
Table~\ref{tab:cl10mass}. The distributions are
satisfyingly symmetrical about the true values; their widths are in 
very
good agreement with the mean of the individual 1-sigma errors calculated from
equation~(\ref{eq:massest}), shown by the solid error bars. 
This demonstrates that the noise 
present
in the galaxy shape data has been successfully carried through to the
inferred quantities. 

We note that if there is no error due to the estimation of the galaxy
shapes (\emph{i.e.} $\sigma_{\rm obs}=0$) then the error bar on the mean
inferred mass is reduced by approximately ten per cent, which corresponds
roughly to the change in the combined ellipticity error~$\sigma$.

\begin{table}
\caption{Mass estimates for CL10. }
\label{tab:cl10mass}
\begin{tabular}{ccccc}
 Aperture & $M_{\rm true}$ &  $\langle M \rangle_{100}$ & $\langle \sigma_M \rangle_{100}$ \ \\ \
  1       & 1.18      & $(1.18 \pm 0.12)$        &     0.12      \\
  2       & 0.39     & $(0.36 \pm 0.07)$       &     0.06     \\
  3       & 0.28     & $(0.27 \pm 0.06)$       &     0.07     \\
          & 0.12     & $(0.14 \pm 0.06)$        &     0.07     \\
\end{tabular}

\medskip
All masses are in units of~$10^{14}
h^{-1}$ M$_{\odot}$. Left to right, the columns contain: aperture number
(see text), true projected mass within aperture, 
the mean and standard
deviation of masses estimated from 70~arcsec reconstructions from 100
noise realisations of the dataset, and the mean inferred error of these
100 mass estimates. The final row contains mass estimation results for
the third aperture translated to a region containing apparently very
little mass.
\end{table}

\begin{figure}

\centering\epsfig{file=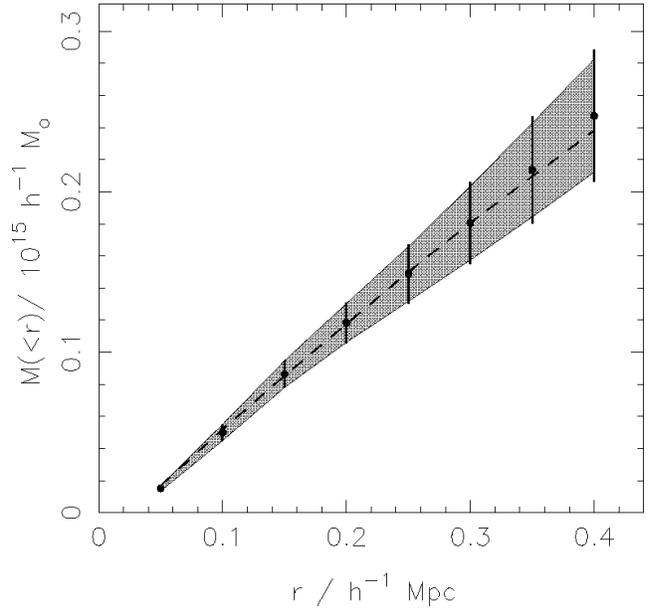,width=0.95\linewidth,angle=270,clip=}

\caption{Mass profile for the largest sub-clump of cluster CL10. 
The points shows the mean estimated
mass within that radius while the error bar is the
inferred 1-sigma error (averaged over 100 noise realisations). 
The shaded area shows the 1-sigma dispersion in the mass estimates over
the 100 realisations.
The dotted line marks the true profile. $0.4 h^{-1}$ Mpc corresponds
to 170~arcsec.}
\label{fig:cl10profile}
\end{figure}

We can take this analysis one step further and calculate a mass profile
around the larger sub-clump; this is shown in
Figure~\ref{fig:cl10profile}. The mass estimates can be seen to be
accurate over a reasonable range of angular scales, with slight
overestimation as the apertures extend further outside the observing 
region.

Our analysis has at no point attempted 
to account for the
`mass sheet degeneracy'~\cite{FGS85,SS95}.
If the 
observing field is sufficiently large then the entropic prior 
acts
to pin down the reconstruction at the edge of this region~(\ppii), 
constraining 
any mass sheet transformation to be small. This control of the mass 
sheet degeneracy is
the outcome of our choice of a low default model value to be used in the
cross-entropy function~(see \ppii). The value used in all
reconstructions in this work was 100~h~$M_{\odot}$pc$^{-2}$; a lower
value was found to leave the reconstruction maps and mass
estimates unaffected. However, significantly increased model values 
gave masses 
overestimated by some tens of per cent, with mass sheets visibly present
in the reconstructions.    
With the default model set suitably low, the residual effect of the mass
sheet degeneracy on the mass estimates
in any given reconstruction is small compared to the uncertainties 
due to the noise realisation; this can be seen by comparing the widths 
of the histograms to the ensemble-averaged error bars in 
Figure~\ref{fig:cl10hist}.

Interestingly, the higher resolution maps
give mass estimates which are systematically below the true
values: such maps provide closer fits to the noise in the data, which
breaks up the coherent lensing signal leading to an underestimate of 
the total mass present. This also illustrates the way in which the value of~$w$
preferred by the evidence is determined by the fit across the whole
image, not just at the peaks of the mass distribution. 
\begin{figure*}
\begin{minipage}[t]{0.48\linewidth}
\centering\epsfig{file=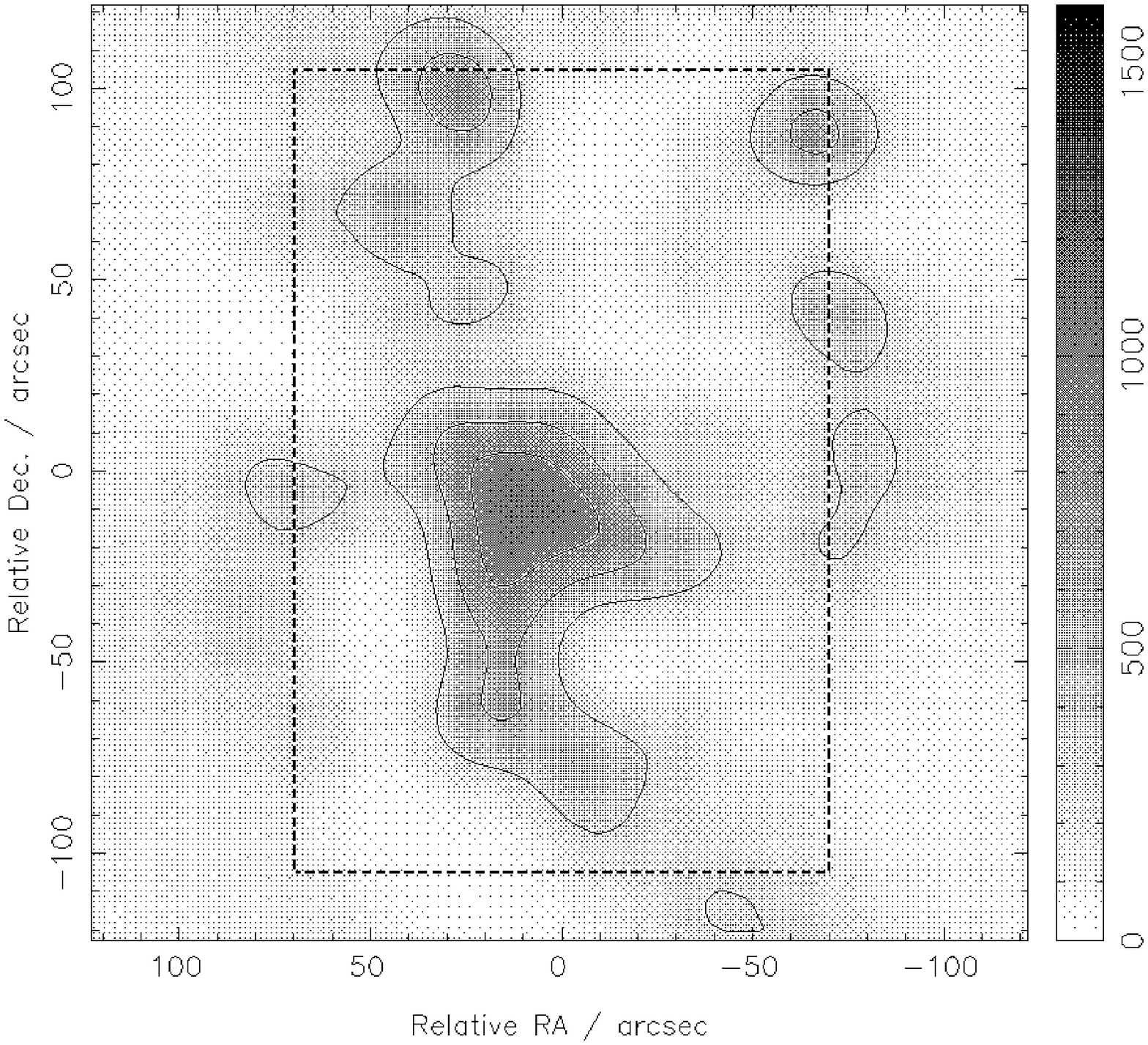,width=\linewidth,angle=0,clip=}
\end{minipage} \hfill
\begin{minipage}[t]{0.48\linewidth}
\centering\epsfig{file=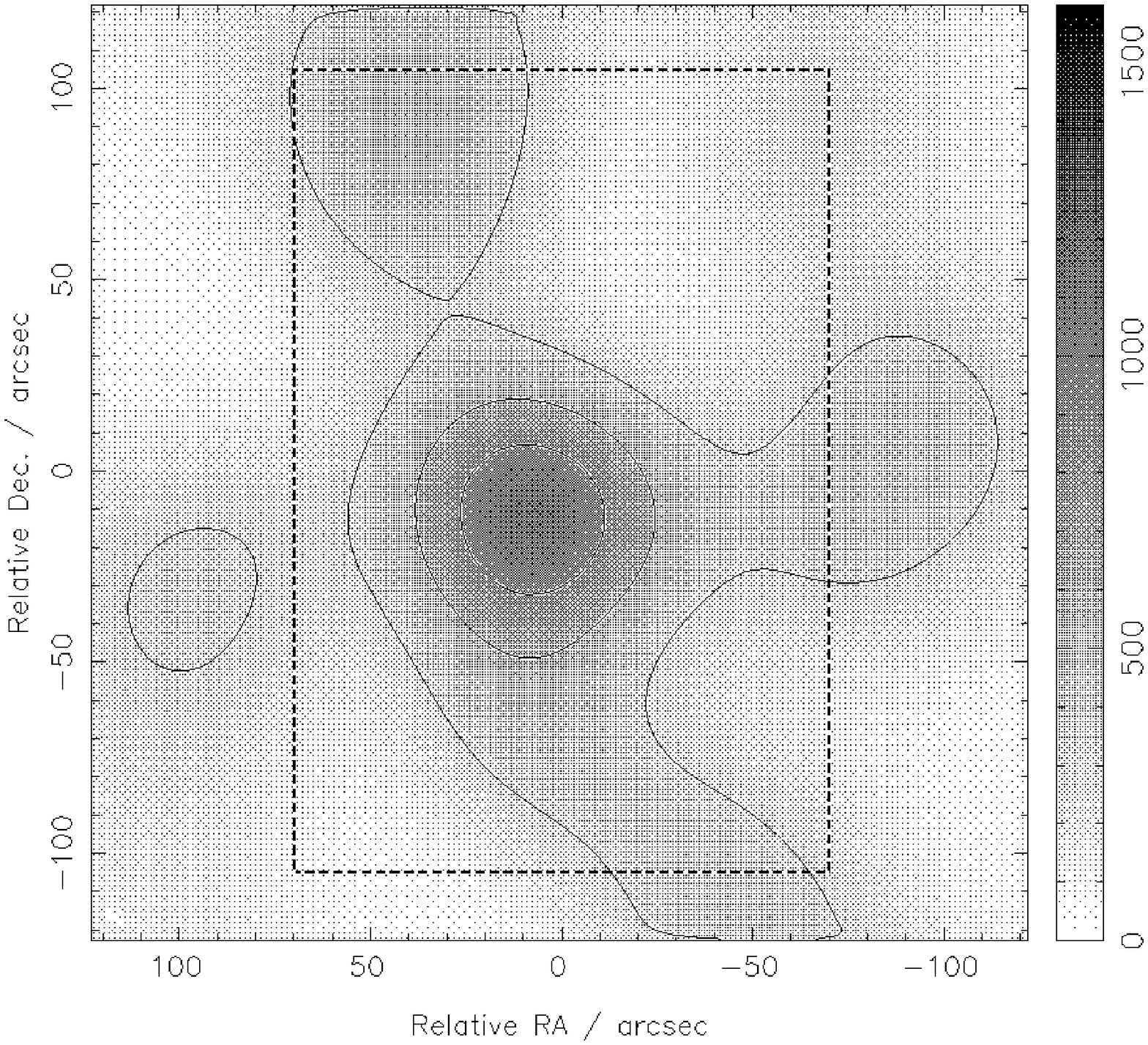,width=\linewidth,angle=0,clip=}
\end{minipage} 
\caption{Top: Reconstructed mass distributions for the cluster CL08.
Left to right, the ICF width parameter~$w$ increases from 20 to
50~arcsec. 
Contours show surface density in steps of 300~h~$M_{\odot}$
pc$^{-2})$.The maximum on the density scale corresponds to a 
convergence of 0.38.}
\label{fig:cl08recon}
\end{figure*}

Although our mass estimation procedure is simplistic, it
does produce sensible and accurate results, even for mass
condensations located on the edges of the observing region.
The filament-like structure
lying between the subclumps of CL10, which was hinted at in the 
reconstruction
maps of Figure~\ref{fig:cl10recon}, was successfully detected in that 
region; its mass was measured with an uncertainty of $\sim 20$~per cent.  
The same aperture was translated to the North-East by approximately 
200~arcsec, to a region containing just 
$0.12 \times 10^{14} h^{-1}$~M$_{\odot}$. 
When the mass estimation analysis was repeated, this value was also
recovered to within the mean inferred error of $\sim 40$~per cent.

\subsubsection{CL08}
\label{cl08}

\begin{figure}
\begin{minipage}[t]{0.48\linewidth}
\centering\epsfig{file=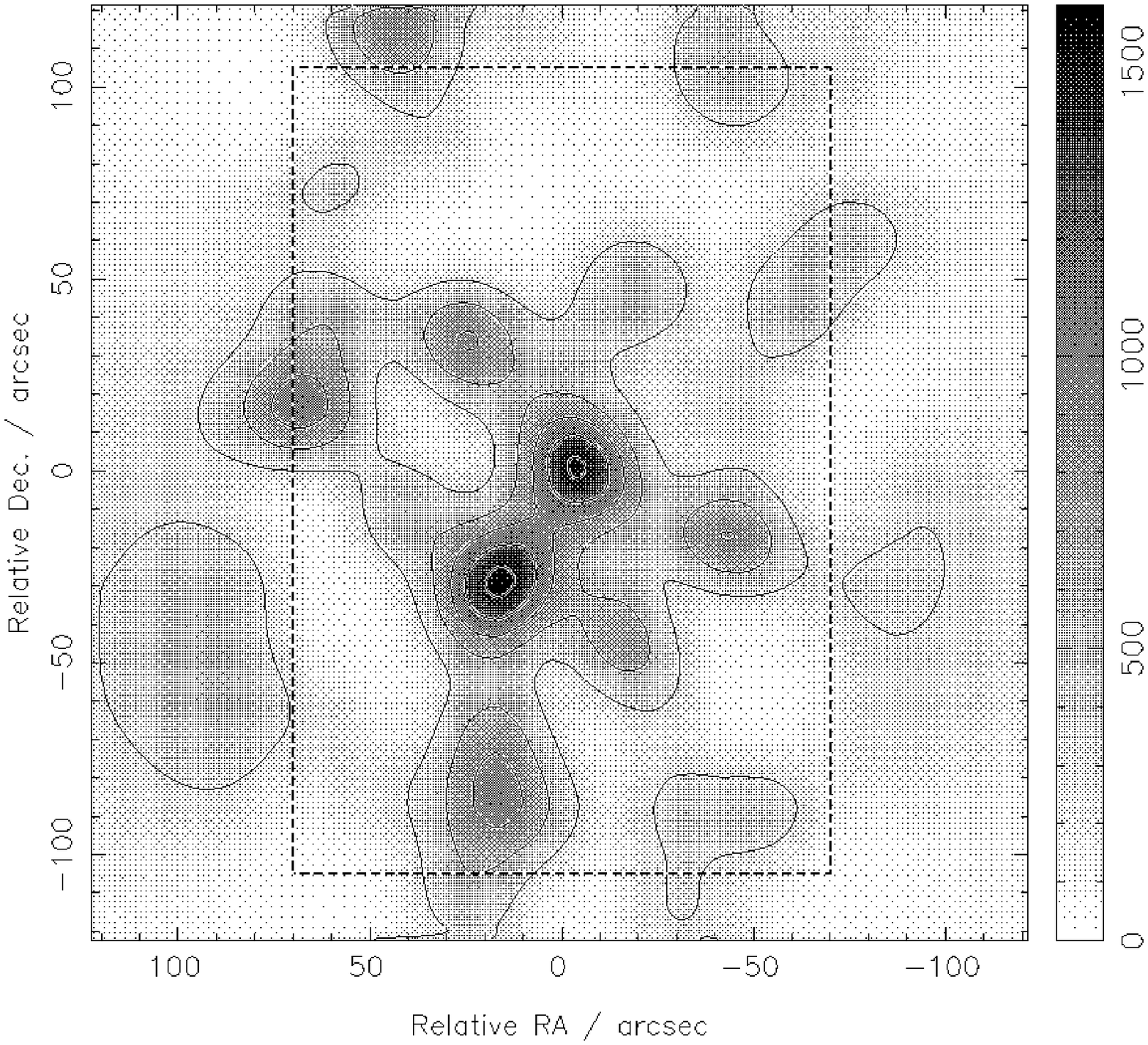,width=\linewidth,angle=0,clip=}
\end{minipage} \hfill
\begin{minipage}[t]{0.48\linewidth}
\centering\epsfig{file=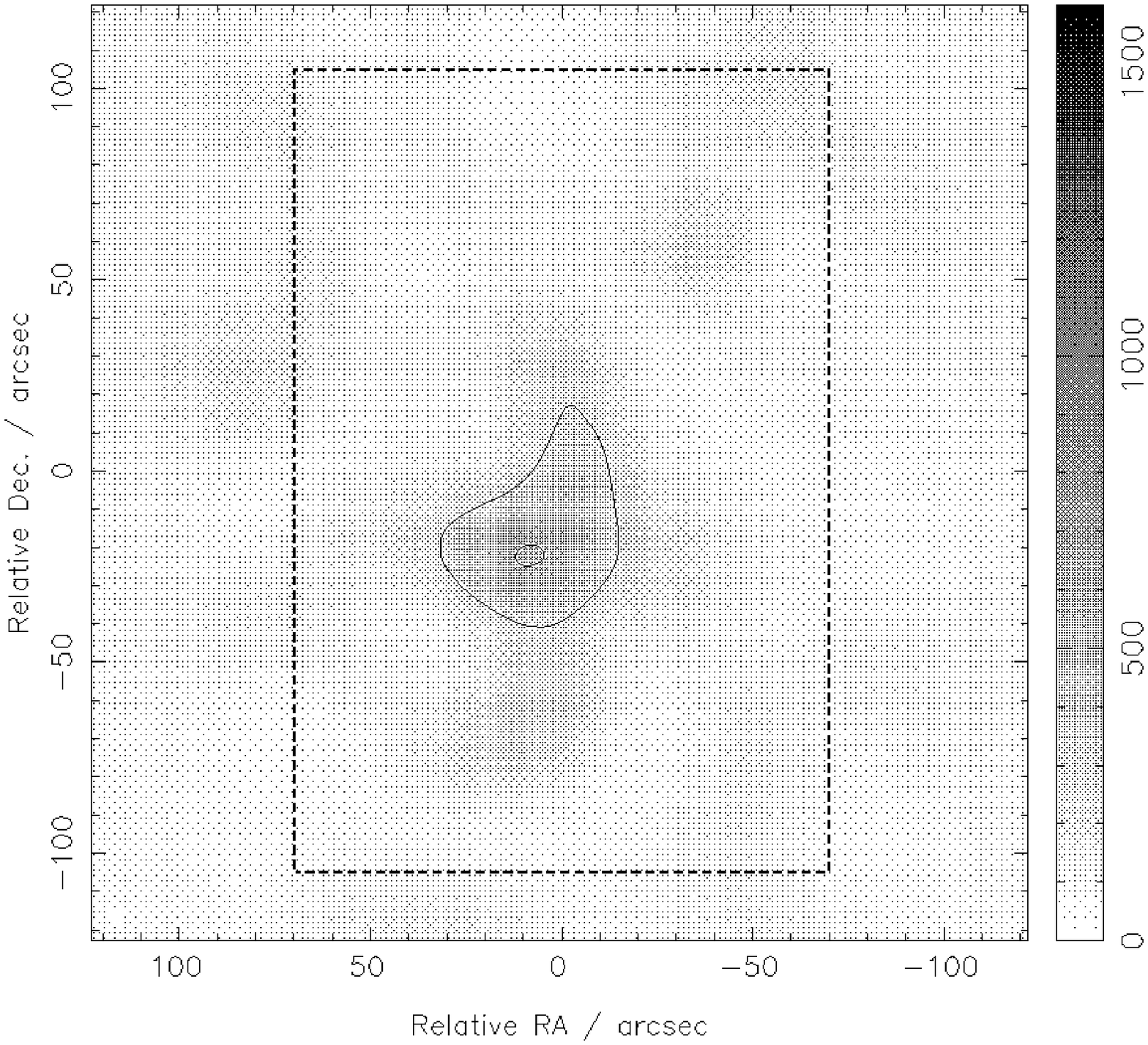,width=\linewidth,angle=0,clip=}
\end{minipage} 
\vspace{3mm}

\begin{minipage}[t]{0.48\linewidth}
\centering\epsfig{file=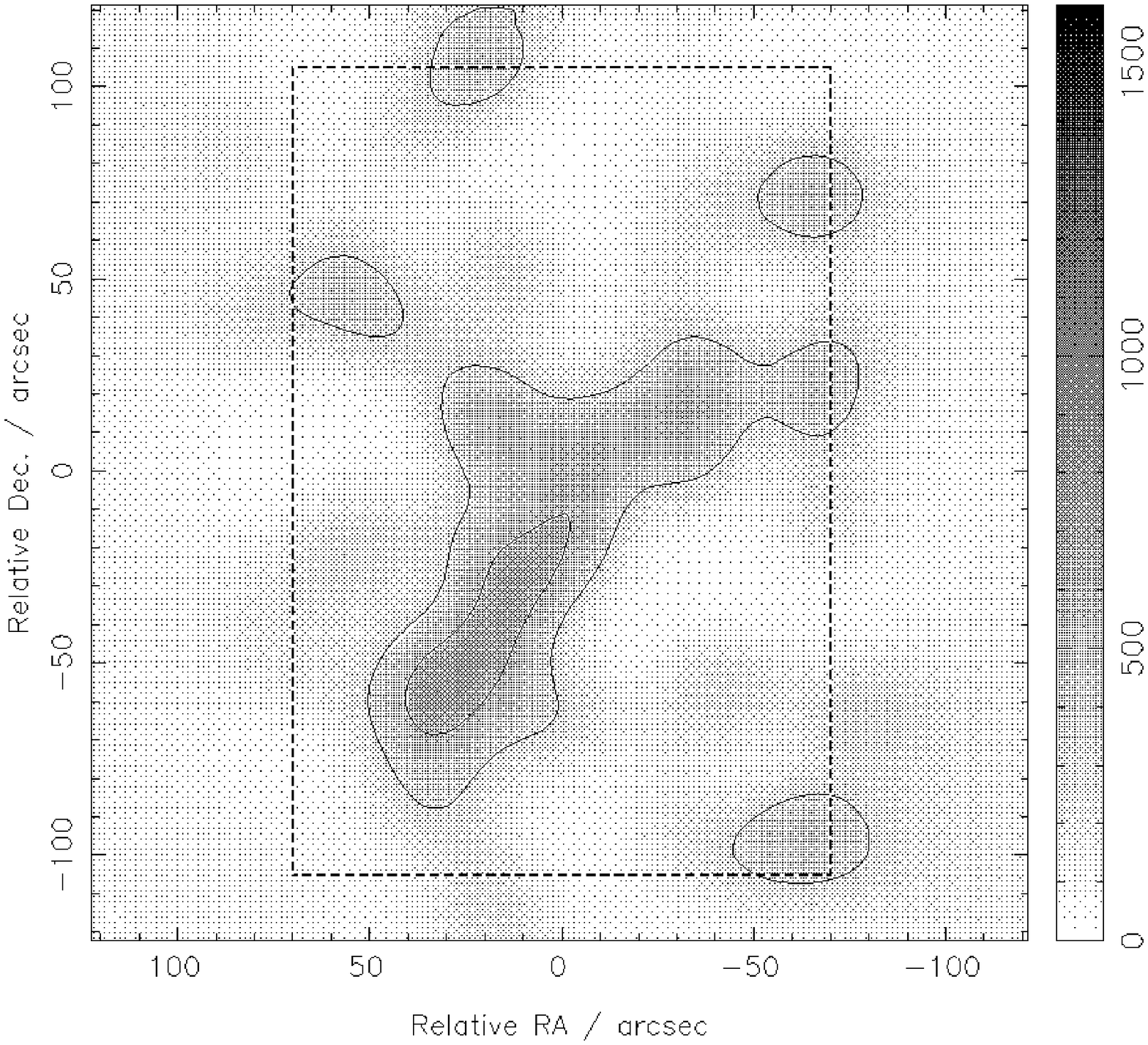,width=\linewidth,angle=0,clip=}
\end{minipage} \hfill
\begin{minipage}[t]{0.48\linewidth}
\centering\epsfig{file=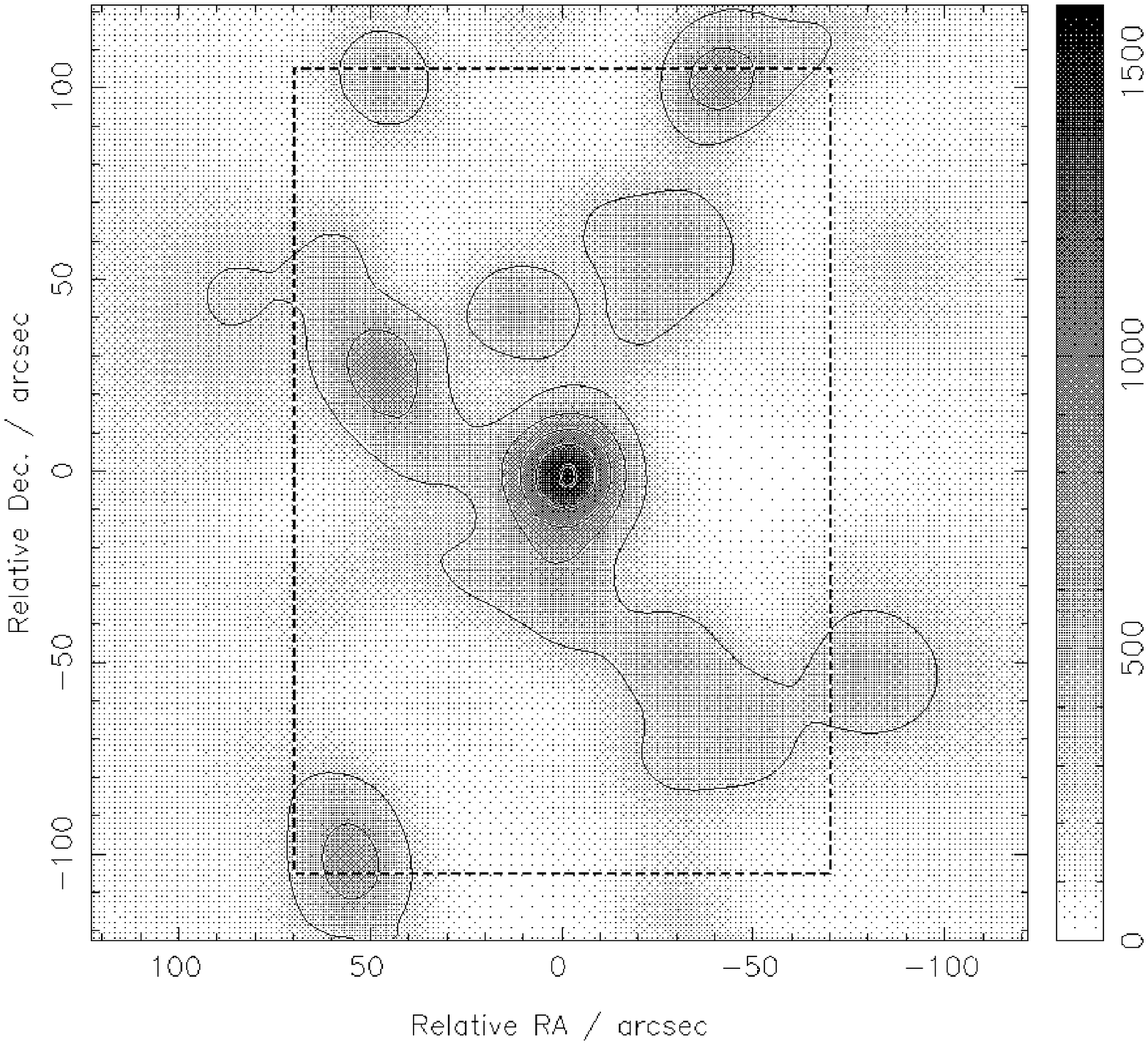,width=\linewidth,angle=0,clip=}
\end{minipage} 
\caption{High resolution ($w=20$~arcsec) reconstructions of CL08,
from 4 different realisations of the background galaxy population.}
\label{fig:cl08multi}
\end{figure}

With a smaller dataset and lower peak convergence, this simulation 
presents a more difficult problem. The most probable Gaussian ICF 
width is
found to be 50~arcsec and the corresponding reconstruction is shown in
Figure~\ref{fig:cl08recon}. The substructure in the cluster has been
smoothed over, and the peak density is underestimated by a 
factor
of two.
Although a higher resolution reconstruction, which is also shown in 
Figure~\ref{fig:cl08recon}, does suggest 
substructure, this particular noise realisation does not enable the 
fine detail of the true
cluster mass distribution to be faithfully recovered. As an
illustration,
20 arcsec ICF reconstructions of four different noise
realisations are
shown in Figure~\ref{fig:cl08multi}. These reconstructions were
deliberately selected (from a larger sample of 20) to highlight the 
extremes of good and bad fortune. 
The presence of density peaks in the reconstructions is clearly
sensitive to the particular noise realisation. In contrast, the 
50~arcsec reconstructions of the noise realisations of 
Figure~\ref{fig:cl08multi} are all very similar.
In practice, there is
only ever one available data set; for the set analysed in 
Figure~\ref{fig:cl08recon} there is very little information about the
two density peaks, but there \emph{is} a lensing signal from the broader
underlying mass distribution.
The 50~arcsec
reconstruction is the most probable given the data: all of the
data are used to infer the global noise properties, which are then 
reflected in the smooth reconstruction. Given such data we can say only
that cluster CL08 is extended in the North-South direction, and that
there is a slight suggestion of substructure.  

\begin{figure}
\centering\epsfig{file=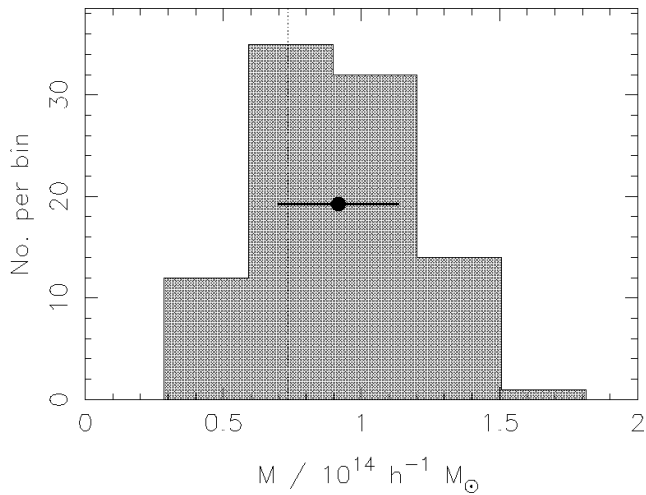,width=\linewidth,angle=0,clip=}
\caption{Mass estimates from 50~arcsec reconstructions of CL08 (see 
Figure~\ref{fig:cl10hist}. 
The aperture, shown in Figure~\ref{fig:cltrue} is a circle of radius 
$0.25 h^{-1} Mpc$.}
\label{fig:cl08hist}
\end{figure}

\begin{figure}

\centering\epsfig{file=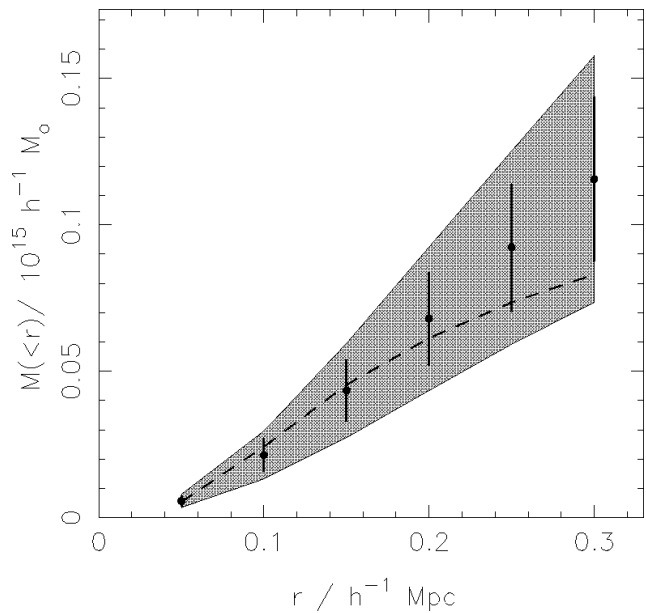,width=0.95\linewidth,angle=270,clip=}

\caption{Mass profile for CL08. 
The points shows the mean estimated
mass within that radius while the error bar is the
inferred 1-sigma error (averaged over 100 noise realisations). 
The shaded area shows the 1-sigma dispersion in the mass estimates over
the 100 realisations.
The dotted line marks the true profile. $0.4 h^{-1}$ Mpc corresponds
to 80~arcsec.}
\label{fig:cl08profile}
\end{figure}

Despite the apparent poor quality reconstruction, the shear data are 
still sensitive to the total mass within an aperture: the
mass estimation histogram of Figure~\ref{fig:cl08hist} shows the total 
projected mass within 0.25~h$^{-1}$Mpc of the cluster centre 
to be well constrained by
the data, with the large error bars (on average 
$0.2 \times 10^{14} h^{-1}$~M$_{\odot}$) comparing
reasonably well 
with
the width of the histogram ($0.3 \times 10^{14} h^{-1}$~M$_{\odot}$). 
This discrepancy reflects the larger
impact of the residual mass sheet degeneracy when smaller observing 
fields are used.
There is a tendency towards overestimation of the total mass
as the additive noise in the reconstruction becomes important for this
less massive cluster,
but this effect is within the estimated error for the aperture
considered
(mean=0.92, truth=$0.74 \times 10^{14} h^{-1}$~M$_{\odot}$).
The effect can be seen more clearly in Figure~\ref{fig:cl08profile}
the corresponding figure to Figure~\ref{fig:cl10profile}. 
Note that the edge of the observing region is at $0.3 h^{-1}$Mpc; in
such low signal, high noise, small observations, extrapolation beyond
the immediate vicinity of the cluster is clearly to be done with care.
One issue is that of the prior on the ICF width -- as the ICF width
approaches the size of the observation the effect of the mass sheet
degeneracy will increase. 

\subsection{More advanced analyses}
\label{advanced}

The above analysis used a circularly symmetric
Gaussian intrinsic correlation function, but there is no 
\emph{a~priori} reason why this smoothing kernel should give the best results.
We also experimented with circular top-hat, exponential and softened
isothermal (`beta') 
profiles, and all were found to give significantly 
lower values of the
evidence for a given data set than the Gaussian ICF. The softened
isothermal profile, aiming to optimise the fit to the cluster
profile, was found to be worse at suppressing the noise in the
outer regions of the cluster, while
the presence of such broad wings introduced a large systematic 
overestimation of the mass integrals due to the mass sheet degeneracy.
It is quite possible that the
optimal ICF for the weak lensing reconstruction problem is not Gaussian,
but our experience with these three alternative functions
leads us to expect that any gain in evidence would be marginal, 
and the reconstruction for a given ICF width would be changed very
little.

The argument given above against using the isothermal profile for an ICF
suggests the use of more than one ICF at a time, 
allowing multiple
resolution scales in the reconstruction. The 
reconstruction then consists
of a weighted sum of convolutions of hidden images with varying
width ICFs. This `multi-scale maximum-entropy' method has been applied 
to a number of problems~\cite{Wei92,Bon++94,Mac++02}; it allows
high spatial resolution where the data warrant it. 
However, when multi-scale ICFs were applied to the weak lensing
problems shown here, very little increase in evidence was found over 
the single ICF reconstructions, and the inferred mass distributions 
from the two approaches were indistinguishable.
The introduction of another hidden image increases the size of the
hypothesis space, introducing extra complexity to the reconstruction
which is not justified by the quality of the data.


\section{Application to real data}
\label{realdata}

\begin{figure*}
\begin{minipage}[t]{0.48\linewidth}
\centering\epsfig{file=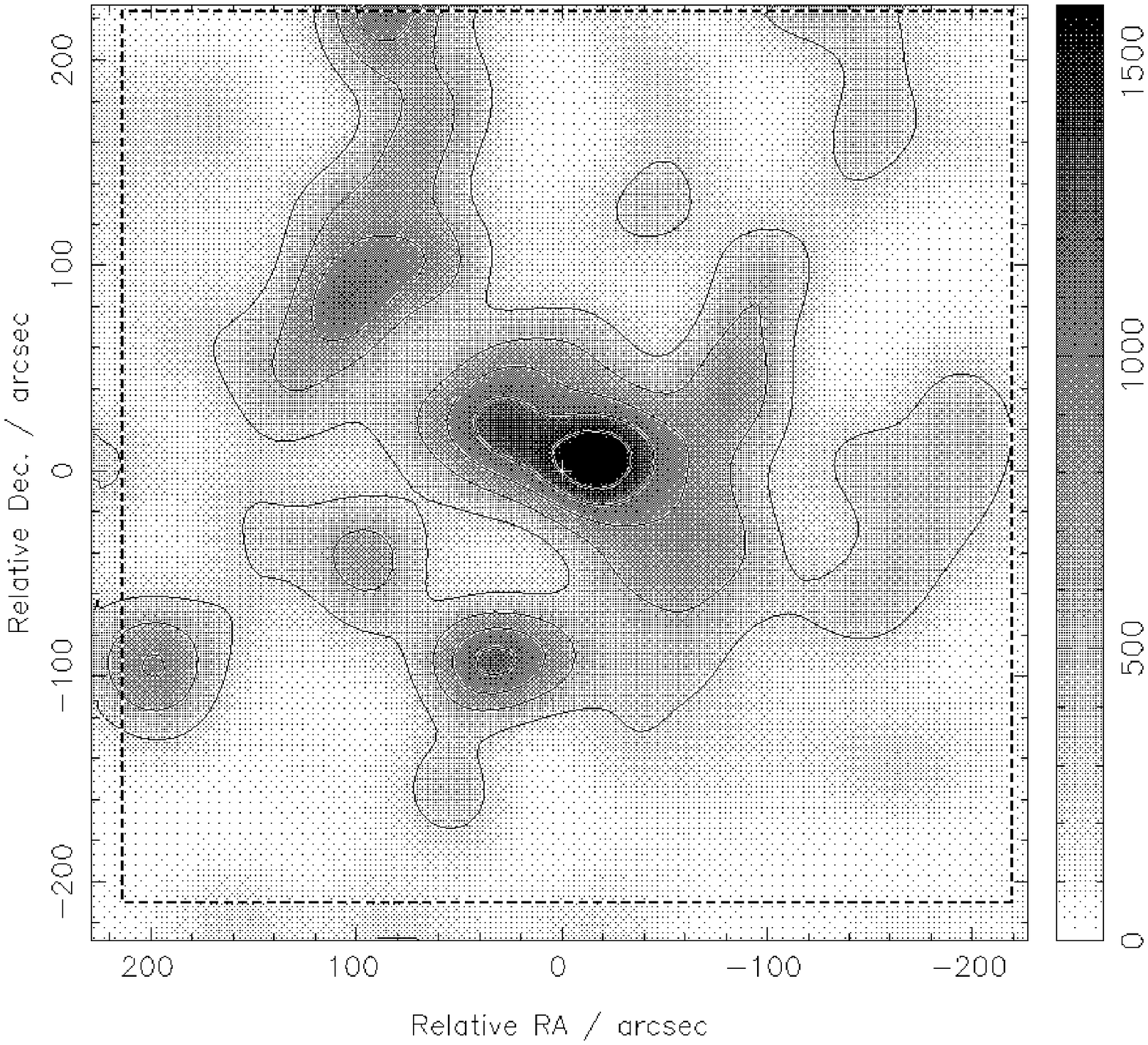,width=\linewidth,angle=0,clip=}
\end{minipage} \hfill
\begin{minipage}[t]{0.48\linewidth}
\centering\epsfig{file=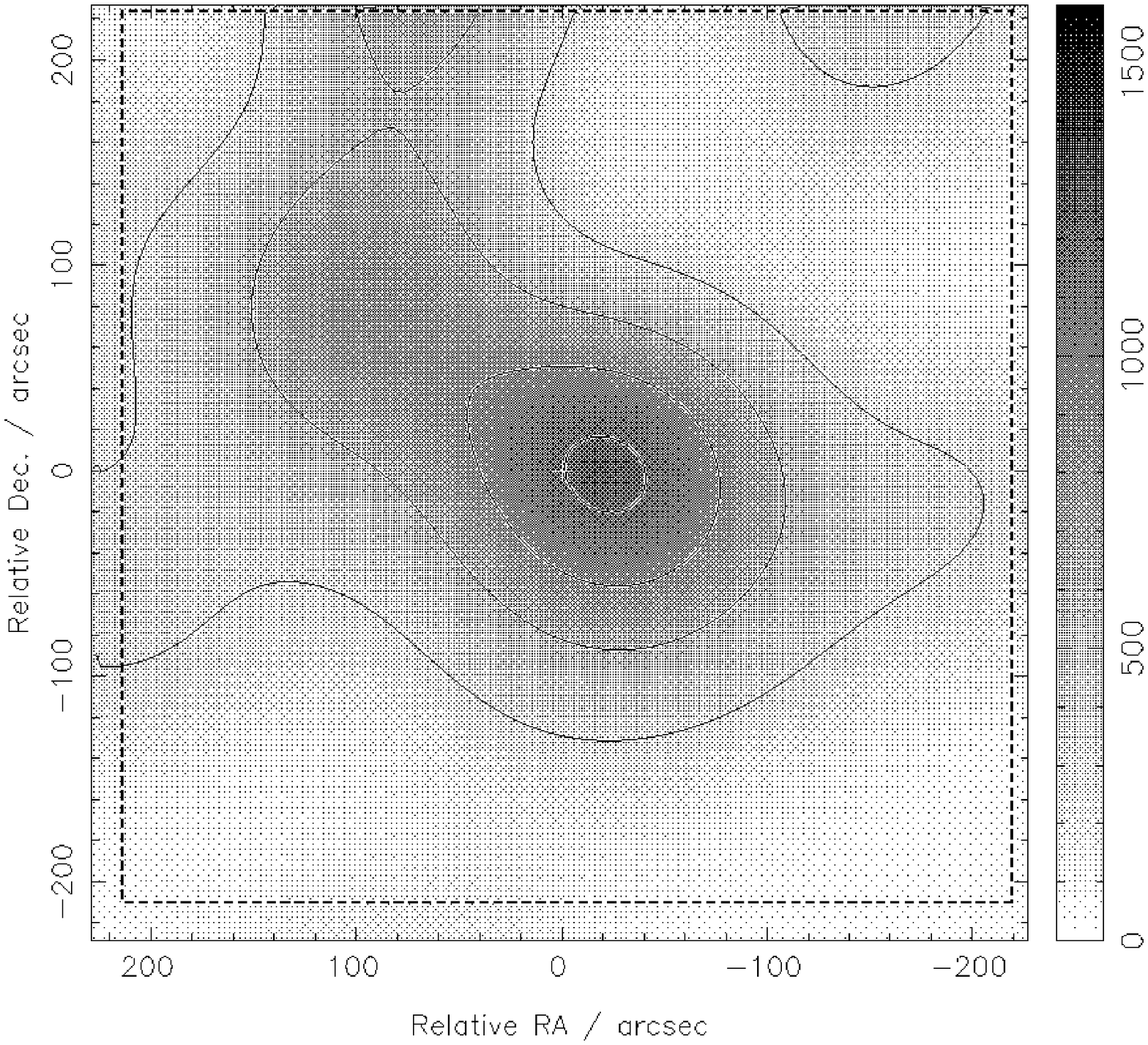,width=\linewidth,angle=0,clip=}
\end{minipage} 
\vspace{5mm}

\begin{minipage}[t]{0.48\linewidth}
\centering\epsfig{file=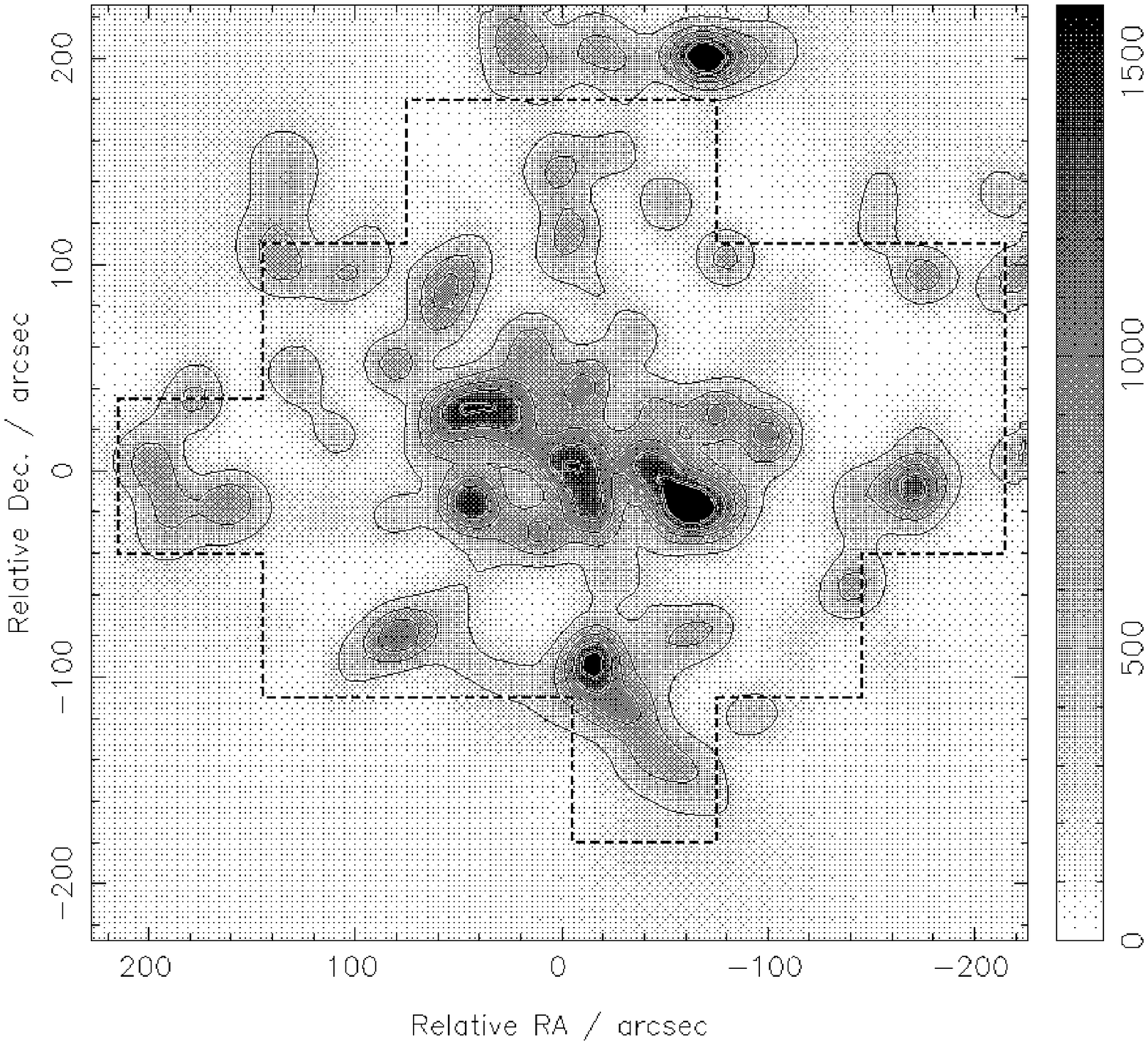,width=\linewidth,angle=0,clip=}
\end{minipage} \hfill
\begin{minipage}[t]{0.48\linewidth}
\centering\epsfig{file=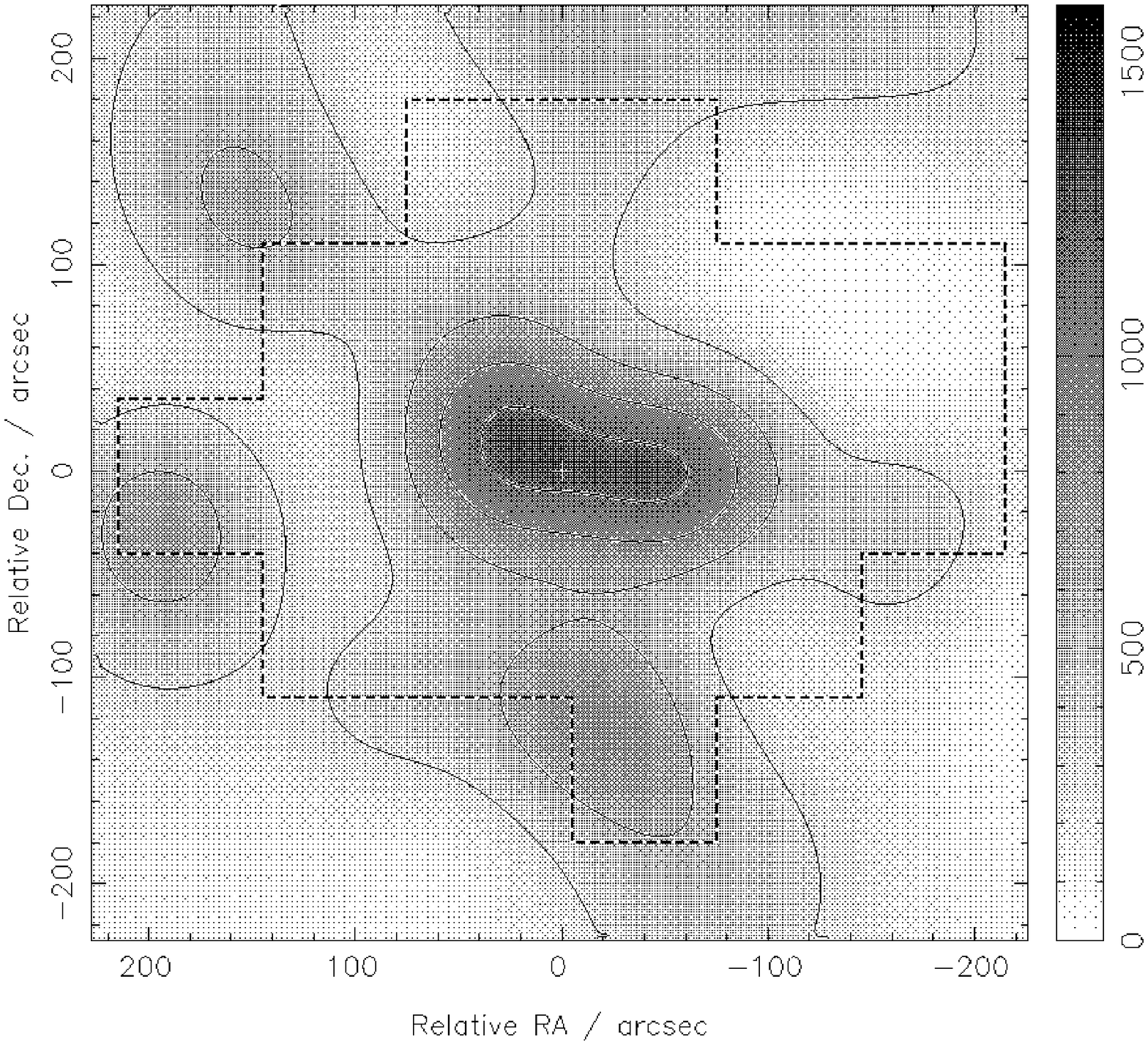,width=\linewidth,angle=0,clip=}
\end{minipage} 
\caption{Top: Reconstructed mass distributions for the cluster 
MS1054-03.
Top: 40~arcsec (left) and 120~arcsec (right) ICF width reconstructions
from the Keck data of Clowe~\ea
Bottom: 20~arcsec (left) and 80~arcsec (right) ICF width reconstructions
from the HST data of Hoekstra~\ea 
The left-hand panels contain maps with angular resolution 
corresponding to that of the maps already published; 
the right-hand panels show the maximum evidence
reconstructions.
Contours show surface density in steps of 300~h~$M_{\odot}$
pc$^{-2}$. The maps are centred on the brightest cluster galaxy (BCG)
position, marked with a cross.} 
\label{fig:ms1054recon} 
\end{figure*}

We now apply our maximum-entropy method to real data.
MS1054-03 is a high redshift ($z=0.83$) galaxy cluster; X-ray
and dynamical measurements suggest that it has a high mass 
($T_X \approx 10 \rm keV$, Jeltema \ea 2001, $\sigma \approx 1150$
km~s$^{-1}$, van Dokkum 1999).  
Two sets of weak lensing data have been analysed.
Clowe~\ea\shortcite{Clo++00} produced a catalogue of 2723 background
galaxies from a single Keck LRIS pointing, with a number density of
approximately 50-60 galaxies per square arcminute. They performed a KS93 
inversion using a smoothing kernel with a FWHM of approximately 
40~arcsec,
and found the cluster to be extended in the East-West direction; with
a smaller smoothing kernel they find three mass peaks to
lower significance. Hoekstra~\ea\shortcite{HFK00}
measured the ellipticities of 2446 galaxy images from a deep HST mosaic
consisting of 6 interlaced WFPC2 fields. They achieved a source density
of
around 80~arcmin$^2$. They used the maximum probability extension to the 
KS93
algorithm~{\cite{SK96} to produce a higher resolution map (smoothed with
a 20~arcsec kernel) showing three
distinct mass peaks.

The maximum entropy reconstructions from these data sets are shown in
Figure~\ref{fig:ms1054recon} for two ICF widths, a low value of~$w$
equal to that used in the previously published analysis, 
and the width that 
maximises~$\pr(w|\mbox{Data})$. All reconstructions were performed using the
Gaussian ICF on a $128\times128$ pixel grid. Hoekstra~\ea calculate
observational uncertainties on their estimated galaxy shape parameters
and add them in quadrature to the intrinsic dispersion; we did the same.
No corresponding weighting of galaxies
is present in the Keck dataset.

\begin{figure}
\centering\epsfig{file=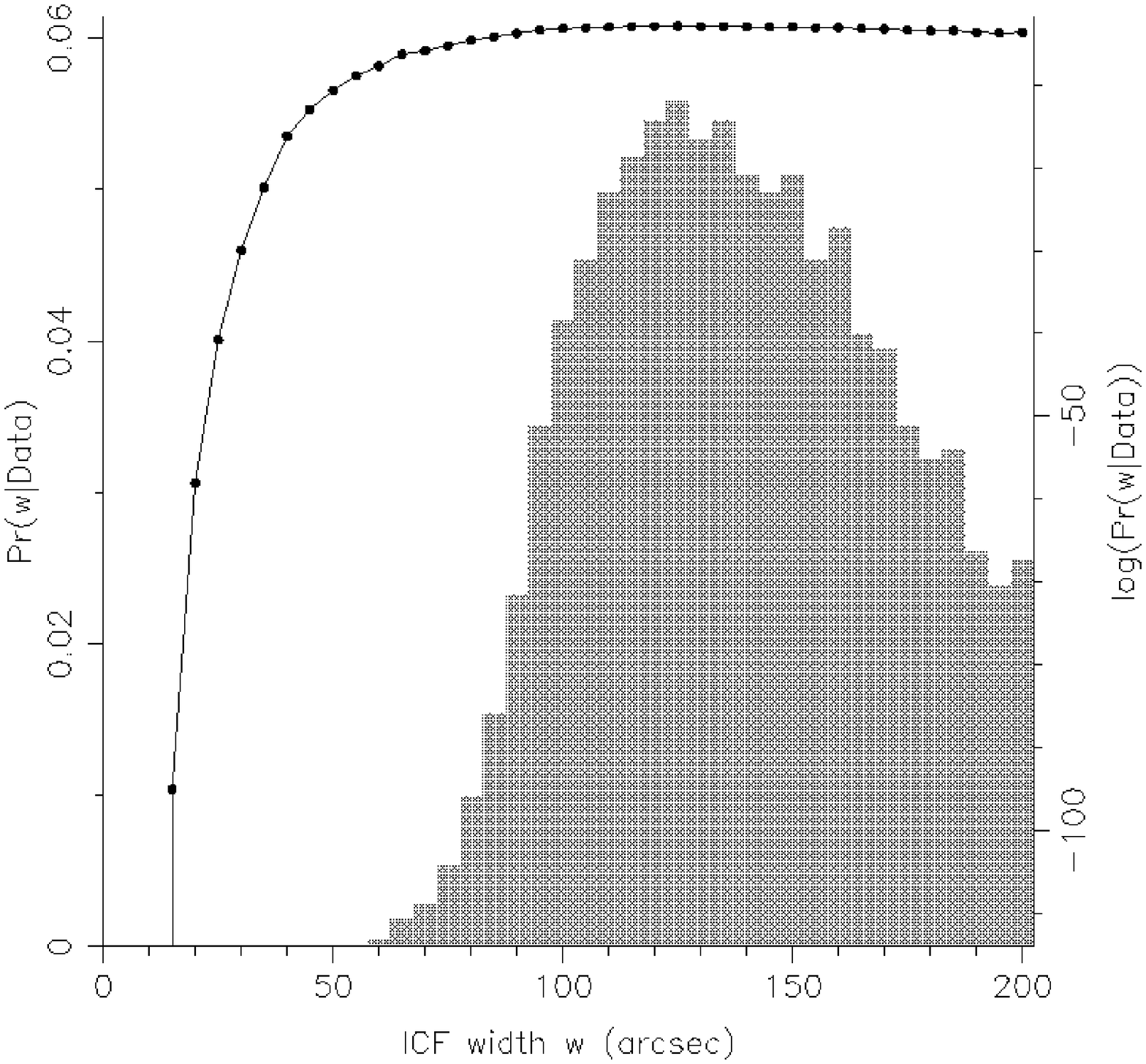,width=0.95\linewidth,angle=0,clip=}
\vspace{3mm}

\centering\epsfig{file=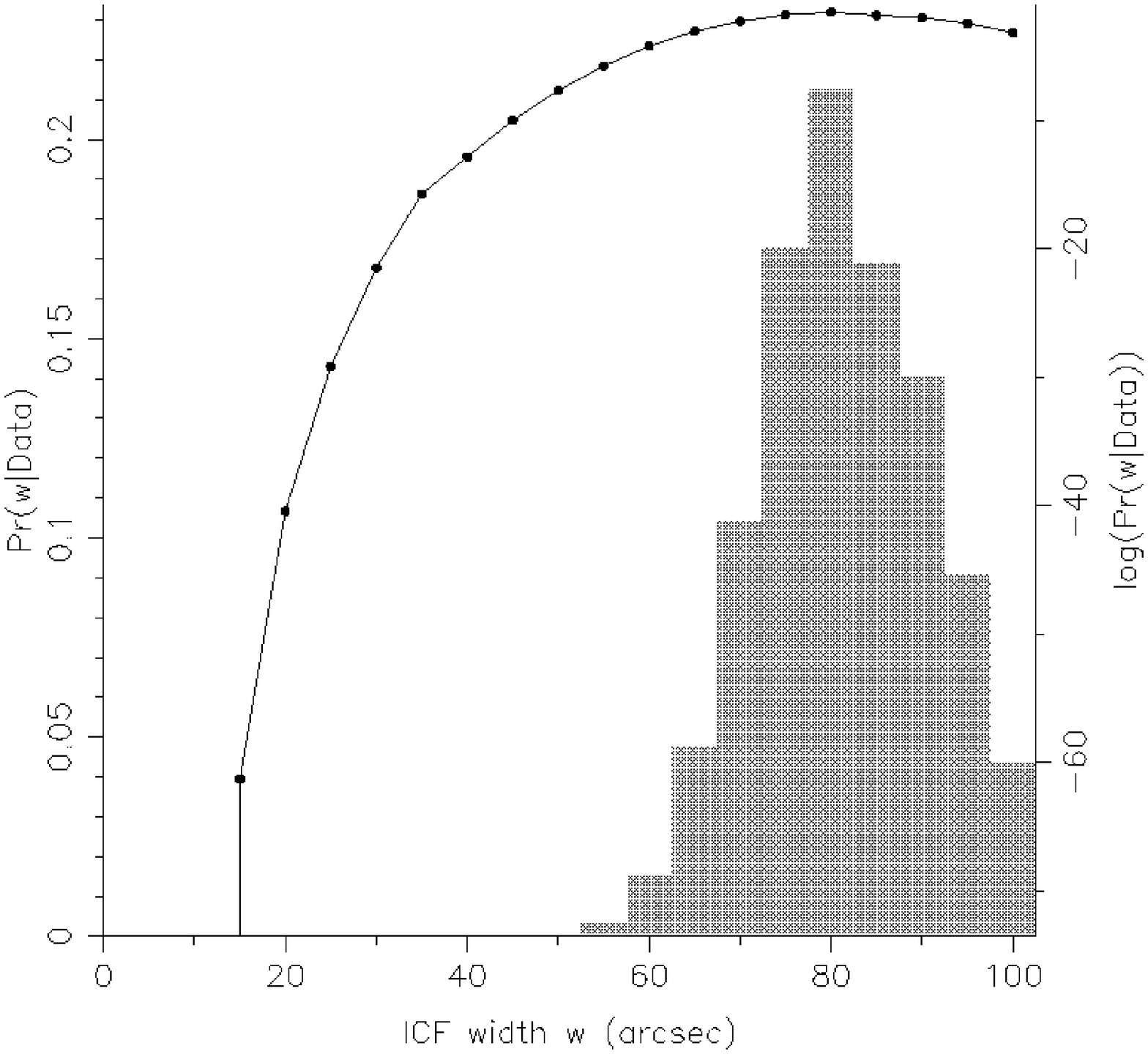,width=0.95\linewidth,angle=0,clip=}
\caption{$\pr(w|\mbox{Data})$ for the MS1054-03 analyses. 
Top: Keck data; bottom: HST data.}
\label{fig:ms1054evid}
\end{figure}

Both of the high resolution maps are qualitatively very similar to those given
in the
referenced papers (where the data were smoothed with Gaussian kernels of
the same width as the chosen low~$w$ ICFs),
but are now positivity-constrained. In particular, the 
three peaks found by Hoekstra~\ea\shortcite{HFK00} are reproduced
along with several
other features present towards the edges of their map.
The reconstruction from the Keck is also similar to the one found by
Clowe~\ea\shortcite{Clo++00}.

We now consider the probability
distribution of the ICF width parameter~$w$. This is shown for the two
datasets in Figure~\ref{fig:ms1054evid}. In both cases the evidence
peaks at a significantly larger value of~$w$ than that used in the high
resolution maps. These `maps of believable features' clearly show 
the absence of significant structure away from the central
cluster region, and suggest that the quality of the data is such
that the substructure observed in the high resolution maps of 
Figure~\ref{fig:ms1054recon} should be interpreted with caution. 
A measure of the goodness-of-fit of the two different resolution maps to
the data was obtained by calculating the chi-squared statistic of
equation~\ref{eq:chisq}, with the summation now running over the galaxies
contained within 40~arcsec radii apertures placed over the three
candidate sub-clumps inferred from the HST data. Reduced chi-squared
values were calculated by dividing by the number of galaxies in the
aperture ($\approx 280$) minus the
number of pixels in the aperture ($\approx 100$); these are given in
table~\ref{tab:chisq}. 

\begin{table}
\caption{Reduced chi-squared values for each of the three candidate 
sub-clumps in MS1054-03.}
\label{tab:chisq}
\begin{tabular}{ccc}
  Sub-clump           & $\chi^2_{20}$ & $\chi^2_{80}$ \ \\ \
   West               &       1.18    &     1.20        \\
   Centre             &       1.09    &     1.10        \\
   East               &       1.20    &     1.26        \\
\end{tabular}

\medskip
The subscripts refer to the resolution scale~$w$ of the reconstruction.
Note that in each case the half-width of the reduced 
chi-squared distribution is approximately $\sqrt{2\times180}/180=0.1$.
\end{table}

It can be seen from this table that there is only a marginal 
improvement
in the fit to the data by decreasing the ICF width; this improvement is
heavily outweighed by the ``Occam's razor'' factor present in the
Bayesian evidence, which suggests that the extra complexity in the
inferred mass
distribution introduced by using an ICF width of less than 80 arcsec is
not justified by these data alone.

The two datasets are by no means independent noise realisations, since
they are both observations of the same background galaxy population.
However, the different observing conditions have clearly introduced
different galaxy shape measurement errors. The resulting 
difference in the details of the high resolution maps of
Figure~\ref{fig:ms1054recon} apparently accords with the conclusions 
drawn from the probability distribution 
of the resolution parameter~$w$. However the differences will also
partly be
due to the different weighting of the images in the two
datasets.

As stated before, in the absence of any other information about the
cluster 
the maximum
evidence map represents the most probable mass distribution given the
data; the
additional information present in the cluster galaxy light and number
density distributions~\cite{HFK00}, and the X-ray surface brightness
maps~\cite{Don++98,Jel++01} are clearly very important in the detailed
interpretation of the weak lensing data, and ideally should be included
in a joint analysis. 

The projected mass within~$0.5 h^{-1}$~Mpc (94~arcsec in the cosmology
of Section~\ref{simobs}) of the brightest cluster 
galaxy
was calculated using the method outlined in Section~\ref{method}. This
was done for both maps given in the lower panels of
Figure~\ref{fig:ms1054recon}, and the results 
compared with that from Hoekstra~\ea\shortcite{HFK00} 
in Table~\ref{tab:ms1054mass}. Both the high and the 
low resolution mass estimates are consistent with the mass derived,
from the tangential shear in circular apertures about the brightest
cluster galaxy, by Hoekstra~\ea It is reassuring to note that the 
statistical errors on these mass estimates agree very well 
with those calculated by aperture mass densitometry elsewhere.
The issue of which
resolution reconstruction, and so which mass estimate, to prefer
may depend on prejudices about the likely level of substructure in
this cluster. Given no other information about MS1054-03 we would
conclude that the quality of the data suggest the 70~arcsec resolution
map as being the more probable, but that there is strong evidence for
substructure in the core of this cluster; in any case 
the weak lensing data
give a projected mass of $10^{15} h^{-1}$~M$_{\odot}$ with a statistical
error of about 10 per cent.

\begin{table}
\caption{Mass estimates for MS1054-03.}
\label{tab:ms1054mass}
\begin{tabular}{cc}
  Reconstruction:     &   $M / 10^{15} h^{-1}$ M$_{\odot}$       \ \\ \
   $w=20$             &  $(0.91 \pm 0.09)$   \\
   $w=80$             &  $(1.21 \pm 0.15)$   \\
   Hoekstra~\ea 2000  &  $(1.07 \pm 0.12)$   \\
\end{tabular}

\medskip
All masses refer to the projected
mass within $0.5 h^{-1}$~Mpc of the BCG (located at the origin) and are
in units of $10^{15} h^{-1}$ M$_{\odot}$. Only the HST data is used
here, to allow direct comparison with Hoekstra~\ea
\end{table}


\section{Conclusions}
\label{conc}

We have developed a Bayesian analysis, based on the
maximum-entropy method of Bridle~\ea(1998, 2001), for inferring the
distribution of mass in clusters of galaxies from weak lensing shear
data. We treat each background galaxy image as a noisy estimator of the
reduced shear field of the cluster, retaining all the 
information about both signal and noise and so allowing the
for high angular resolution.  
Use of an `intrinsic correlation function' in the
maximum-entropy formalism provides a way of incorporating our prior 
expectation of
clusters as smooth, extended objects, and effectively
replaces the data smoothing required by direct reconstruction methods.
In contrast to the these methods, the lensing signal is not diluted by
this process.
Moreover, analysis of the posterior probability distribution of the 
ICF width~$w$, obtained by numerically evaluating the Bayesian evidence,
provides an objective way of discriminating between 
smoothing
scales. The map at the peak of this probability distribution was found
not
to contain any significant spurious peaks, and can be 
interpreted as the safest conclusion to draw from the data.
The higher resolution maps,
although representing an overfit to the data, do contain limited useful
information particularly with respect to substructure in the cluster;
the fact that their angular resolution is less favoured by the data
quality gives a useful indication of the believability of these
features. 

Simple mass estimates extracted directly from 
the mass maps preferred by the evidence 
were found to be unbiased and accurate to within the estimated errors 
over a fairly wide range of angular scales; 
these uncertainties agreed very well with the standard
deviation in the mass estimates from 100 different realisations of the
background galaxy population. Noisier observations over a smaller field
of view were found to give mass estimates with a slight bias towards
overestimation, an effect understood in terms of the additivity of the
reconstructed distribution and the mass sheet degeneracy. 
Inspection of the variety of structure in
reconstructions from these different noise realisations justified
the cautious interpretation of the high resolution maps, indicating that
this analysis provides a useful way of understanding the noise
properties of the data.

We have applied our method to two galaxy shape datasets for the high
redshift cluster MS1054-03, one derived from a ground based observation
and the other from an HST mosaic. In both cases the features found in 
previously published maps, obtained by both direct and inverse methods,
are reproduced, but with the added desirable features of being
positivity-constrained and 
quantitatively useful. Simple mass estimates extracted directly from 
the reconstruction agree well with values found by aperture mass
densitometry, as do the errors estimated by both methods.

The principal function of parameter-free mass maps as
produced by this method is to provide the equivalent of a mass
telescope, allowing images to be generated to aid further, more
quantitative analysis. Parameterised
fitting to shear data with physical motivation has been performed 
elsewhere~\cite{SKE00,KS01}; we
believe that the information provided by our variable angular resolution
analysis is a helpful guide to this process, whilst also providing
reasonably accurate `mass photometry' at the same time. The use of the
Bayesian evidence in such fitting will be addressed in future
papers.  
  

\subsection*{Acknowledgments}

We thank Henk Hoekstra and Douglas Clowe and their
co-workers for supplying us with their data, and Vincent Eke for 
providing the
simulated projected mass distributions.

Thanks are also due to Andy Taylor, Richard Saunders and 
Charles McLachlan 
for helpful discussions, and to Anton Garret for proof-reading the
manuscript. We also thank the anonymous referee for comments leading to
improvements in the paper.

PJM acknowledges the PPARC for support in the form of a Research 
Studentship. The MEMSYS4 code was developed by Maximum Entropy Data 
Consultants Ltd., and is available for use by the astronomical community
under the name `LensEnt2'. This can be obtained from
\texttt{http://www.mrao.cam.ac.uk/projects/lensent/}


\bsp 
\label{lastpage}
\end{document}